\documentclass[aps,prb,reprint,superscriptaddress,amsmath,amssymb,floatfix]{revtex4-2}

\usepackage{graphicx}
\usepackage{bm}
\usepackage{booktabs}
\usepackage[colorlinks=true,linkcolor=blue,citecolor=blue,urlcolor=blue]{hyperref}
\usepackage{braket}
\usepackage{xcolor}

\graphicspath{{figure/}}

\begin{document}

\title{Conformality lost in Anderson localization in infinite dimensions, \\and regained in finite dimensions: \\a phenomenological theory of the Anderson transition in high dimensions}

\author{Antonello Scardicchio}
\affiliation{The Abdus Salam International Center for Theoretical Physics, Strada Costiera 11, 34151 Trieste, Italy}
\affiliation{INFN, Sezione di Trieste, via Fabio Severo, 34128, Trieste, Italy}
\author{Carlo Vanoni}
\affiliation{Department of Physics, Princeton University, Princeton, New Jersey, 08544, USA}

\date{\today}

\begin{abstract}
We show that the recent works on Anderson localization when the number of spatial dimensions $d\to\infty$ can be framed into the scenario dubbed {\it conformality lost}~\cite{Kaplan2009}, which occurs when two fixed points of the RG collide and become complex. This framing makes it natural for a {\it two-parameter scaling} flow {\it à la} Berezinskii--Kosterlitz--Thouless (BKT) to control all the observables of the theory as conjectured in Ref.~\cite{Vanoni2024}. We also use this as a starting point for an $\epsilon$ expansion of the RG equations in $\epsilon=1/d\ll 1$, and find phenomenological RG equations which fit the numerical results in dimensions $d\gtrsim 4$.

\end{abstract}

\maketitle

\section{Introduction}
\label{sec:intro}

Anderson localization~\cite{Anderson1958absence} is getting close to its 70th birthday, and many things are now understood. A proof of the existence of a localized phase at high disorder exists~\cite{Frohlich83} for regular lattices; a general phenomenology of the critical point valid for all space dimensions exists~\cite{Evers08} which shows multifractality of the wave functions to become more pronounced when $d$ grows; field theory methods have been put forward, both in the replicated and in the supersymmetric framework~\cite{ZirnbauerNPB1986,Efetov1983}, which allow, in principle, exact computations of the observables (provided the field theory is amenable to some perturbative treatment, which it has resisted so far); the nonlinear $\sigma$-model of Wegner and Efetov~\cite{Wegner1979,Efetov1983} substantiated near two dimensions the one-parameter scaling hypothesis of Ref.~\cite{AALR1979}, and an expansion in $d-2$ of the critical exponent~\cite{Wegner1979} was done and carried over to four loops~\cite{hikami1981anderson,hikami1992localization}.

What remains to be done? Certainly, one question that is receiving a lot of attention, due to recent experimental and theoretical advances, is what happens to a localized state of non-interacting electrons when the interaction between them is restored. Studies of this question fall under the umbrella name of many-body localization. Even though the question exists in the work of Anderson already~\cite{Anderson1958absence}, and in Fleishman and Anderson~\cite{FleishmanAnderson1980}, the most recent revival follows the breakthrough work of Basko, Aleiner, and Altshuler (BAA)~\cite{Basko06}. Many researchers~\cite{Pal10,Oganesyan07,Abanin18,sierant2025many} have built on BAA to describe a physical state in which strong disorder hinders all kinds of transport, except entanglement, ending up in a phase with local (or quasi-local) integrals of motion~\cite{Imbrie17,ros2015integrals}.

Leaving interactions aside, the most relevant and pressing question in the mind of a theorist (maybe not so much of an experimentalist, though) is how much mileage one can get from the field theory description. Can we characterize the critical conformal field theory in some way? Can we obtain the critical exponents in some series expansion? The natural expansion in $d-2$ unfortunately does not help. Although the series can be written down to four loops, the results for $d=3$ do not match the numerical results $\nu=1.57...$, not even to the first significant digit. 
Ueoka and Slevin have tried Pad\'e-Borel resummation of this series and compared it to $d>2$ numerical data, but the work is not definitive~\cite{ueoka2014dimensional,ueoka2017borel}. 
The agreement with data is poor, and one does not get much insight from the whole procedure. This is not surprising, since most likely there is no finite upper critical dimension in the Anderson model~\cite{MF94}, and with increasing $d$ the critical point moves to strong disorder, where the perturbative treatment of the $\sigma$-model is no longer controlled~\cite{zirnbauer2023wegner}, and the one-parameter scaling hypothesis itself has been questioned~\cite{Goldenfeld2006}; the expansion around $d=2$ most probably does not capture the physics of high dimensions. 
Alternatively, in recent years, the possibility of an expansion in $\epsilon=1/d$ has been put forward, and therefore the limit of large dimension of the Anderson model has been reconsidered seriously by various groups (including the present authors)~\cite{TM2019,Tarquini17, zirnbauer2023wegner,arenz2023wegner,Vanoni2024,Altshuler2025}. Zirnbauer and Arenz~\cite{zirnbauer2023wegner,arenz2023wegner}, starting from the Wegner $N$-orbital model at strong disorder, have argued that in high dimension the $\sigma$-model must be replaced by a field theory with two independent couplings, two stiffnesses on the hyperbolic target space. On the experimental side, arrangements of cold atoms in effectively four-dimensional lattices have been used to explore the transition~\cite{madani2025observation}. In three dimensions, the mobility edge has also been related to a percolation transition of the localization landscape~\cite{Filoche2024}. A linear fit of the inverse of the numerical exponents in $d=3$--$6$ gives $1/\nu\simeq2-4.75/d$~\cite{Tarquini17}, equivalent to $\nu=\tfrac{1}{2}+\nu_1/d+O(d^{-2})$ with $\nu_1\simeq1.2$.
Along a separate line of research, Parisi and collaborators have used the $M$-layer construction~\cite{altieri2017loop,angelini2024bethe}, where the parameter $M\to\infty$ turns the finite-$d$ lattice into a Bethe lattice, a $1/M$ expansion is done, and then $M$ is set to 1. This route is also promising, but the problem has not been solved yet~\cite{baroni2024corrections}.

Much of the recent interest in the Anderson transition on tree-like graphs came from many-body localization. A non-ergodic extended phase, once proposed~\cite{KAI2018}, has since been ruled out: on these graphs the eigenstates are ergodic at any $W<W_c$, and the apparent non-ergodicity is a finite-size crossover~\cite{TMS2016,BHT2022,GarciaMata2022,sierant2022universality}. What these studies did establish is that the finite-size flow near the transition is not of the one-parameter type: Garc\'{\i}a-Mata et al.~\cite{GarciaMata2022} described the localized side with a two-parameter flow of BKT type, and Sierant et al.~\cite{sierant2022universality} showed that the critical point is localized, which essentially forces a scaling of BKT type.

When we started this program a few years ago, the first step was to reinterpret the classic Bethe lattice results and cast the finite-$N$ flow into an RG flow topology. We showed that the exact diagonalization (ED) numerics~\cite{sierant2022universality,Vanoni2024} is well described by a {\it two-parameter scaling} flow {\it à la} BKT (recently found also on small-world random graphs~\cite{garciamata2026connecting}). The program then was to consider modifications of this flow at order $\epsilon=1/d$, morphing the two-parameter flow into a one-parameter flow. 
If enough of the two-parameter theory (the Bethe lattice) and the deformation (the operator that gives $1/d$) is known, one can hope to obtain the coefficient $\nu_1$ of the previous paragraph analytically and therefore hope for a theory in $1/d$. This paper is a small step in the program.

In this paper, we show that the exact same results that were obtained with ED can be obtained with population dynamics and that the discrepancies that exist in the literature are due to imprecise fits and/or misinterpretation of the theory. In particular, we find that all the $D_q$ flow in the same way under the RG close to the critical point, and that this flow is BKT-like in the system size $N$, with $D_q\propto1/(\ln N)^2$ at the critical point (Section~\ref{sec:popdyn}). We also show that the one-parameter part of the flow, the arc towards the ergodic point sketched in Ref.~\cite{Vanoni2024}, begins where the linearized self-consistency equations break down, and their nonlinear part takes over (Section~\ref{sec:linear}). Related results, including the BKT form of the flow, the value of $\beta$ on the localized line of fixed points and the continuation of the flow to the delocalized side, were obtained independently and simultaneously by Rizzo, Tarzia and Vecchiarelli~\cite{RTV2026lett,RTV2026prb}, from the finite-size scaling of the inverse participation ratio.

Then, we analyze the theory of the fixed point of the iteration equations and show how the transition, which is the merging of a stable and an unstable fixed point which are real at $W>W_c$ and become complex when $W<W_c$, is described by the {\it conformality lost} scenario of Ref.~\cite{Kaplan2009}. The linearized equations near the localized or critical region can be written as a differential equation for the generating function of the imaginary part of the Green's function. Its solutions are power laws whose two exponents, the infrared and ultraviolet fixed points, merge at the transition; in the supersymmetric language, this is a scalar field on AdS$_2$ crossing the Breitenlohner--Freedman bound. Away from the stationary limit, the exponent runs with the scale, and the two variables of Ref.~\cite{Vanoni2024} acquire a precise meaning: $\beta$ is the running coupling and $D_1$ its beta function (Section~\ref{sec:linear}).

Building on this picture, which identifies the RG couplings in the running exponent and the distance from the merger, in the second part of the paper we start an expansion in $1/d$ (Section~\ref{sec:largeD}). This part is of a different nature: we do not derive the corrections from a microscopic expansion, but write the minimal deformation of the tree flow allowed by dimensional counting and by the effects of the loops, with two coefficients that we fix by comparison with the numerical exponents. We show the effects of dimensional counting, which already gives $\nu=\frac12$, in line with other kinds of discussions that start from the Bethe lattice (like $1/M$ expansion). On the tree, the disorder only labels the trajectories; on the lattice, it acquires a scaling dimension and becomes the relevant coupling. We also identify the contribution of the loops of the lattice as the ones that make the exponent irrelevant, split the merged fixed points, and shift $\nu$ to values close to $O(\epsilon)$, to get closer to the numerical ones. Once the two fixed points are split, the critical point is an ordinary fixed point, with one relevant and one irrelevant direction, and the logarithmic corrections of the tree become power laws. Its ultraviolet partner, a fully unstable fixed point, survives at finite $d$, but it lies in a region of the plane of couplings that the physical flow, which starts from the bare couplings of the lattice, cannot enter. It is therefore never observed directly, although it shapes the flow close to the critical point. We also show how to write, in this scheme, RG equations for the observables like the fractal dimensions whose critical values depend on the definition. The final result is that the topology of the flow at finite $d$ is sketched in Fig.~\ref{fig:schematic}, the values of the coefficients of the $1/d$ corrections are conjectured, and we leave their calculation for future work.

\section{Model and population dynamics}
\label{sec:popdyn}

We consider the Anderson model, defined by the Hamiltonian
\begin{equation}
    H = -\sum_{i,j} A_{ij} \ket{i}\bra{j} + \sum_i V_i \ket{i}\bra{i},
\end{equation}
where $A_{ij}$ is the adjacency matrix of the graph under consideration, which can be a random regular graph (RRG) or a Bethe lattice with branching number $K$ (coordination $K+1$), or a $d$-dimensional hypercubic lattice. The i.i.d. random variables $V_i$ are chosen uniformly in the interval $[-W/2,W/2]$. We will use $ K=2$ throughout and consider the middle of the energy band at $E=0$.

On locally tree-like graphs, such as the random regular graph and the Bethe lattice, the local resolvent can be computed with the cavity method. Removing a site $j$ splits the tree into disconnected branches, and the cavity Green's function $G_{i\to j}$ is the diagonal element at site $i$ of the resolvent $(E-i\eta-H^{(j)})^{-1}$ of the graph with $j$ removed. Since the hopping enters only through its square, it satisfies
\begin{equation}
\label{eq:cavity_warmup}
    G_{i\to j}= \frac{1}{E-i\eta-V_i- \sum_{a\in\partial i\setminus j} G_{a\to i} },
\end{equation}
where $\partial i$ denotes the $K+1$ neighbors of $i$. 
The diagonal Green's function is obtained with the full coordination $K+1$,
\begin{equation}
\label{eq:cavity_measure}
    G_{ii}= \frac{1}{ E-i\eta-V_i- \sum_{j\in\partial i}G_{j\to i}}.
\end{equation}
With this sign convention $\operatorname{Im}G_{ii}>0$, and $I_i=\operatorname{Im}G_{ii}=\pi\rho_i(E)$ is the local density of states (LDoS) at site $i$, broadened by $\eta$. We solve Eq.~\eqref{eq:cavity_warmup} by population dynamics (Section~\ref{sec:pd-method}). The distribution of cavity Green's functions is represented by a population of $N$ complex numbers. Each generation builds every new member from $K$ members drawn at random and a fresh on-site energy, and the distribution of $I$ is sampled from Eq.~\eqref{eq:cavity_measure}.
We test the flow on the stationary cavity problem of the Bethe lattice, using
the full complex recursion at finite regularization $\eta$ rather than its
linearization.

\subsection{Method}
\label{sec:pd-method}

\paragraph{Finite $\eta$ as finite size.}
A random regular graph of $N$ sites is locally a Bethe lattice, and 
at the band center $E=0$, the average density of states per site is finite and nonzero~\cite{baroni2024corrections}. The mean level spacing on an RRG of $N$ sites is therefore $\Delta_N \sim N^{-1}$, so using the level spacing as a regulator, with the cavity broadening $\eta$ proportional to $\Delta_N$, gives $\eta\propto N^{-1}$. This identification is not general: it relies on a finite and smooth density of states at the energy considered, and it must be reconsidered when a symmetry makes that energy special, as at the band center of models with chiral symmetry~\cite{vanoni2026semifractality}.
We therefore employ $\ell=\ln(1/\eta)$ as RG time, with an additive constant left free. With $I=\operatorname{Im}G$ the moments scale as $\langle I^q\rangle_\eta\propto\eta^{(1-q)(1-D_q)}$. 
Since $\langle I\rangle=\pi\rho(E)$, with $\rho(E)=\langle\rho_i(E)\rangle$ being the mean density of states, is smooth at the band center and therefore independent of $\eta$, differentiation at $q=1$ gives
\begin{equation}
  \begin{gathered}
  S(\ell)\equiv\frac{\langle I\ln I\rangle}{\langle I\rangle}=(1-D_1)\,\ell+\text{const},\\
  D_1(\ell)=1-\frac{dS}{d\ell},
  \qquad
  \beta=\frac{d\ln D_1}{d\ell}.
  \end{gathered}
  \label{eq:pd-D1}
\end{equation}
This is the Shannon-entropy definition of Ref.~\cite{Vanoni2024}, with the replacements $|\psi|^2\to I\eta$ and $\ln N\to\ln(1/\eta)$. The identification of $\ln(1/\eta)$ with $\ln N$, with $\eta$ of the order of the mean level spacing, is standard~\cite{MirlinFyodorov1991,TM2019}. That the finite-$\eta$ and finite-$N$ corrections share the same exponent, which is what a comparison of the two flows requires, was shown in Ref.~\cite{RTV2026lett}. Their amplitudes can differ~\cite{RTV2026lett}, so the coefficient $c$ of the flow in $N$ need not coincide with the one measured at finite $\eta$.

\paragraph{Algorithm.}
We iterate the cavity recursion in Eq.~\eqref{eq:cavity_warmup}, which we now write in a more compact form as $G_i=(E-i\eta-V_i-\sum_{j=1}^{K}G_j)^{-1}$, on a population of $N$ complex numbers. 
Each generation replaces the whole population, with the $K$ children chosen at random. 
Measurements use the site Green's function in Eq.~\eqref{eq:cavity_measure}, with $K+1$ neighbors.

In the localized and critical regimes, the $I$-weighted averages in Eq.~\eqref{eq:pd-D1} are dominated by the tail of the LDoS distribution $P(I)$, which extends up to $I\sim1/\eta$ and comes from the site's own resonance. These resonances occur with probability \(\sim\eta\), so naive sampling would resolve them only for \(N\gg1/\eta\). A related difficulty in the computation of IPRs in the localized phase was circumvented in Ref.~\cite{rizzo2024localized} by treating the rare on-site resonances analytically. Here we use a finite-\(\eta\) version of this idea, integrating the energy of the measured site out exactly. 

Denoting $Z=E-i\eta-\sum_{j=1}^{K+1}G_j$, $v=\operatorname{Re}Z-V$ and $b=-\operatorname{Im}Z>0$, one has $I=b/(v^2+b^2)$; the substitution $v=b\tan\theta$ gives $I\,dV=d\theta$, so that, for any test function $f(I)$,
\begin{equation}
  \begin{gathered}
  \int\frac{dV}{W}\,I\,f(I)
  =\frac{\theta_2-\theta_1}{W}\,
   \Bigl\langle f\bigl(\cos^2\theta/b\bigr)\Bigr\rangle_{\theta\in[\theta_1,\theta_2]},\\
  \theta_{1,2}=\arctan\frac{\operatorname{Re}Z\mp W/2}{b},
  \end{gathered}
\end{equation}
with $\theta$ uniform. 
Each member of the population thus carries a pair $(I,w)$, with $I=\cos^2\theta/b$ for a single $\theta$ drawn uniformly in $[\theta_1,\theta_2]$ and $w=(\theta_2-\theta_1)/W$. The weight $w$ is the exact integral $\int dV\,I/W$ at fixed neighbors, and it stays bounded, $w<\pi/W$, as $\eta\to0$. The $I$-weighted averages are then $\langle I\,f(I)\rangle=\langle w\,f(\cos^2\theta/b)\rangle$, where the average on the right runs over the population and over $\theta$. The resonance of the measured site therefore no longer limits the statistical error.
As a check, $\langle I\rangle$ is independent of $\eta$ to better than $10^{-3}$ over eight decades.

\subsection{Numerical results}
\label{sec:pd-results}

\begin{figure*}[t]
\centering
\includegraphics[width=0.9\textwidth]{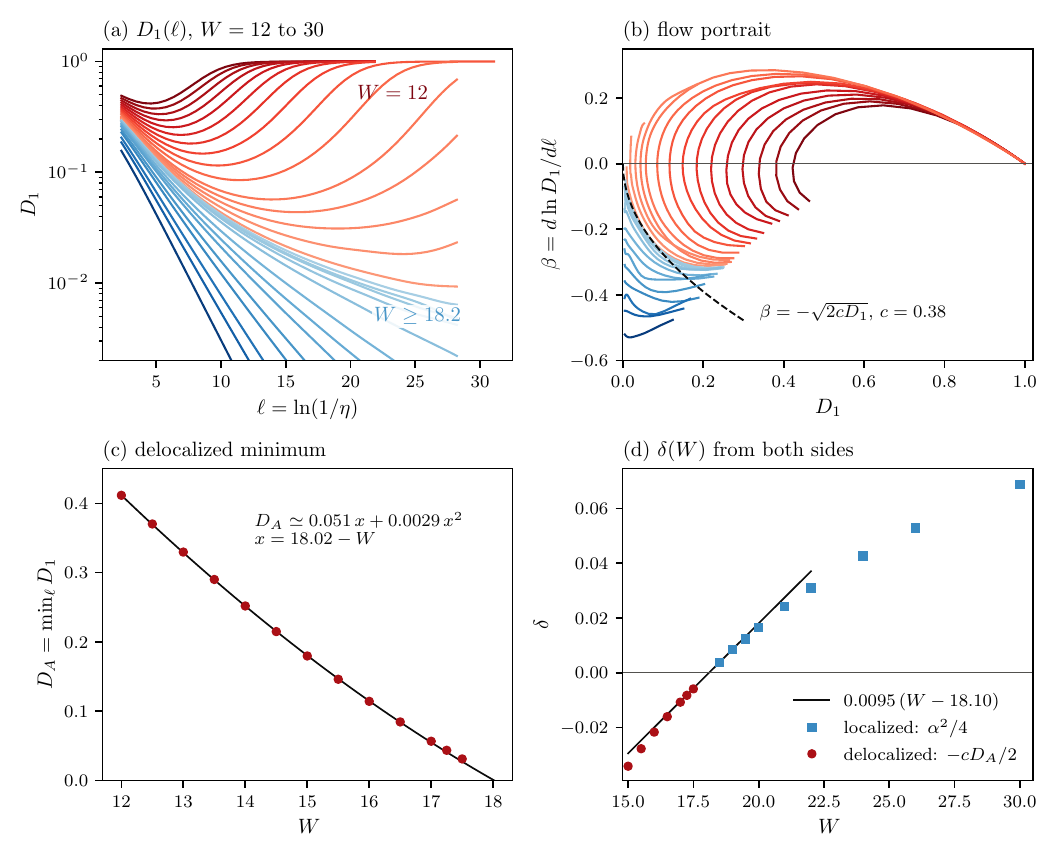}
\caption{Population dynamics on the Bethe lattice ($K=2$, $E=0$, $N=10^7$),
from 100 independent populations evolved with common random numbers in
$\eta$ (Appendix~\ref{app:popdyn}). (a)~$D_1(\ell)$ for 28 values of $W$
from 12 to 30 (red: delocalized, $W<W_c$; blue: localized, $W>W_c$; darker away from $W_c$), from local quadratic fits of $S(\ell)$ over five
points; points with an error larger than $0.02$ are omitted. (b)~The flow
in the $(D_1,\beta)$ plane, from seven-point fits; points with an error on
$\beta$ larger than $0.03$, or with $D_1<0.004$, are omitted. Dashed: the
critical separatrix $\beta=-\sqrt{2cD_1}$ with $c=0.38$. (c)~The
delocalized minimum $D_A(W)$, with the quadratic guide
$D_A\simeq0.051\,x+0.0029\,x^2$, $x=18.02-W$. (d)~$\delta(W)$, the parameter of Eq.~\eqref{eq:mupm}, from the
localized decay rate ($\alpha^2/4$) and from the delocalized minimum
($-cD_A/2$), with the line $\delta=0.0095\,(W-18.10)$.}
\label{fig:pd-prod}
\end{figure*}

Figure~\ref{fig:pd-prod} summarizes the flow of $D_1$, obtained from 100
independent populations of $N=10^7$ members at 28 disorders between $W=12$
and $30$, and for $\eta$ down to about $10^{-13}$. The runs, the error
analysis and the fits are described in Appendix~\ref{app:popdyn}. Three
regimes are visible in Fig.~\ref{fig:pd-prod}(a). On the delocalized side,
$D_1$ first decreases, reaches a minimum $D_A$ and then rises to the ergodic
value $D_1=1$. Near $W\simeq18.1$ it decays as a power of $\ell$. On the
localized side it decays exponentially, at a rate that increases with $W$.
In the $(D_1,\beta)$ plane [Fig.~\ref{fig:pd-prod}(b)] the delocalized
trajectories cross $\beta=0$ at $D_A$ and return to $\beta=0$ at $D_1=1$:
this is the arc found in Ref.~\cite{Vanoni2024} on random regular graphs,
obtained here on the loopless Bethe lattice. The localized trajectories run
towards $D_1=0$ with $\beta<0$, and the two families are separated by the
critical trajectory $\beta=-\sqrt{2cD_1}$.

These are the features of the two-parameter flow derived in
Section~\ref{sec:lin-flow}, Eq.~\eqref{eq:betaflow}, in which a single
relevant parameter $\delta$ changes sign at $W_c$. Integrated, the flow gives
$D_1(\ell)$ on both sides of the transition in terms of $c$, an offset
$\ell_0$ and $\delta(W)$. A common fit of the disorders closest to the
transition gives $W_c\simeq18.1$, $c\simeq0.4$, $\ell_0$ of order one and
$\delta$ linear in $W-W_c$. At $W_c$ the flow predicts
$D_1=2/[c(\ell+\ell_0)^2]$, and the data follow this law up to the largest
$\ell$ reached, with $\ell=\ln(1/\eta)$ entering at unit coefficient. Away from
$W_c$ the same constants account for both sides of the transition. On the
delocalized side the minimum $D_A=2|\delta|/c$ vanishes linearly as
$W\to W_c$ [Fig.~\ref{fig:pd-prod}(c)], and its position grows as
$(W_c-W)^{-1/2}$: this is the walking regime. On the localized side the
decay rate $\alpha=2\sqrt\delta$ of $D_1$ vanishes as $(W-W_c)^{1/2}$, while the
localization length has $\nu_{\rm loc}=1$~\cite{MirlinFyodorov1991}. The values of $\delta(W)$ obtained
independently from the two sides fall close to a single line through $W_c$
[Fig.~\ref{fig:pd-prod}(d)]. Near the ergodic point the flow is linear and the slope is $(d-2)/d\to 1$, as shown in \cite{Vanoni2024}.

The description is quantitative near the localization transition and
degrades away from it, where $D_A$ is no longer small or the localized decay
is fast. There the data resolve corrections to the leading-order flow, and
$c$, estimated from the curvature at the delocalized minimum, decreases
along the arc away from $W_c$.

\begin{figure}[t]
\centering
\includegraphics[width=\columnwidth]{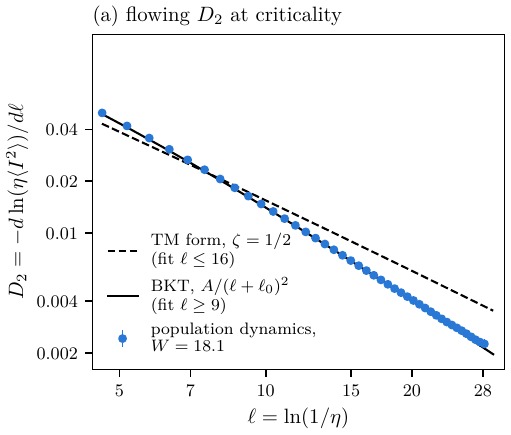}
\includegraphics[width=\columnwidth]{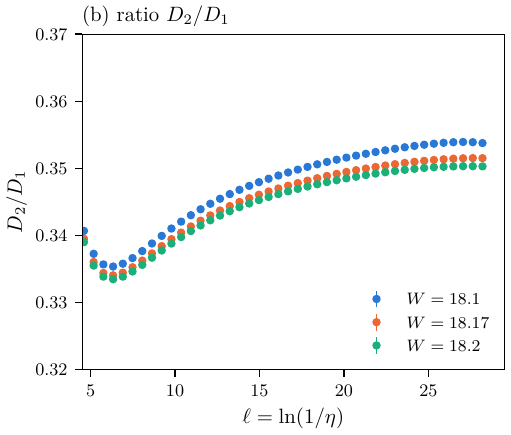}
\caption{(a) Flowing $D_2=-d\ln(\eta\langle I^2\rangle)/d\ell$ at $W=18.1$, from
local quadratic fits over five points of the same data as
Fig.~\ref{fig:pd-prod}, with the derivative of
$\eta K_1=c_1+c_2\ln^{-1/2}(1/\eta)$~\cite{TM2019} fitted on $\ell\le16$
(dashed) and of $\ln(\eta\langle I^2\rangle)=\text{const}+A/(\ell+\ell_0)$ fitted on
$9\le\ell\le29.4$ (solid). (b) $D_2/D_1$ at $W=18.1$, $18.17$ and $18.2$. Error
bars are smaller than the symbols.}
\label{fig:pd-D2}
\end{figure}

The same runs give $\eta\langle I^2\rangle$, and hence the flowing
$D_2=-d\ln(\eta\langle I^2\rangle)/d\ell$ (Fig.~\ref{fig:pd-D2}). With the site
energy integrated out, $\eta K_1(0)\equiv\eta\langle|G_{ii}|^2\rangle=2\eta\langle I^2\rangle+O(\eta)$
is the quantity studied in Ref.~\cite{TM2019}, where the critical behavior
$\eta K_1=c_1+c_2\ln^{-\zeta}(1/\eta)$ with
$\zeta\simeq\frac12$ was proposed. The flow predicts instead
$D_2\propto(\ell+\ell_0)^{-2}$, i.e.\ $\zeta=1$ with an offset. Our raw data
agree with those of Ref.~\cite{TM2019}, but their derivative does not:
$\zeta=\frac12$ fails to describe $D_2$ at $W=18.1$, while
$A/(\ell+\ell_0)^2$ describes it over the whole range
[Fig.~\ref{fig:pd-D2}(a)]. Fitted over a finite window, a form without
offset returns an effective exponent $\zeta_{\rm eff}=\ell/(\ell+\ell_0)<1$,
which grows with $\ell$. The flow also predicts that all $D_q$ follow the
trajectory of $D_1$ with a $q$-dependent amplitude,
$D_q=r_q\,D_1[1+O(m-\mu_+)]$. Near the transition, the ratio $D_2/D_1$ is indeed
almost constant [Fig.~\ref{fig:pd-D2}(b)], with a slow drift that reflects
the $O(1/\ell)$ correction. It is also constant along the localized
trajectories, while on the delocalized arc it rises towards one, as it must
since all $D_q\to1$ at the ergodic point (Appendix~\ref{app:pd-D2}).

Arenz and Zirnbauer~\cite{arenz2026wegner} have recently predicted, for the Wegner one-orbital model in the self-consistent approximation that is exact on a tree (where it reduces to the equation solved here), a singular continuous critical spectrum with $\langle|G_{ii}|^2\rangle\propto\eta^{-1/2}$, hence $D_2=\frac12$ at $W_c$. Our data exclude this. Moreover, $D_2\le D_1$, and we show that $D_1\to0$ at $W_c$. We suspect that the difference comes from their double-scaling limit, in which the states of pure-point type might end up receiving a vanishing weight; at fixed $\eta$, the case relevant for $N\sim1/\eta$, a finite fraction of the cavity Green's functions keeps $I<\eta$.

\section{Linearization, a field theory, and the RG flow}
\label{sec:linear}
In this section we study the linearized problem. Its stationary solutions are power laws in the variable conjugate to $\operatorname{Im}G$, with two exponents that merge at the transition and become complex in the {\it delocalized} phase (Section~\ref{sec:lin-merger}), as in the scenario of Kaplan et al.~\cite{Kaplan2009}. The linear problem is that of a free scalar field on AdS$_2$ (Section~\ref{sec:lin-ads}), and the two exponents are the infrared (IR) and ultraviolet (UV) fixed points of an RG flow (Section~\ref{sec:lin-flow}).

Consider the recursion equations for the cavity propagator,
Eq.~\eqref{eq:cavity_warmup}, broken into real and imaginary parts $G_i=R_i+i I_i$, where $R_i\in\mathbb{R},\ I_i\in\mathbb{R}_+$, so that $G_i$ lives in the complex upper half-plane. Separating the two parts, one gets
\begin{equation}
  \begin{gathered}
  R_i=\frac{E-V_i-\sum_jR_j}{|Z_i|^2},\qquad
  I_i=\frac{\eta+\sum_jI_j}{|Z_i|^2},\\
  |Z_i|^2=\Bigl(E-V_i-\sum_jR_j\Bigr)^2+\Bigl(\eta+\sum_jI_j\Bigr)^2,
  \end{gathered}
  \label{eq:RI}
\end{equation}
with $j$ running over the $K$ children of $i$.

\subsection{The localized solution and its linearization}
\label{sec:lin-loc}

At $\eta=0$, $I_i=0$ on every site is an exact solution of
Eq.~\eqref{eq:RI}. The real parts then obey the closed recursion
$R_i=1/(E-V_i-\sum_jR_j)$, which has a stationary distribution at any
$W$. This is the localized solution: the local density of states
$\rho_i=I_i/\pi$ vanishes at every site. The transition is the point where
it becomes unstable, where a small imaginary part, created by an
infinitesimal $\eta$, no longer dies out as it propagates up the tree. To
first order in $\eta$ and in the $I_j$,
\begin{equation}
  R_i=\frac{1}{E-V_i-\sum_{j}R_j},\qquad
  I_i=R_i^2\Bigl(\eta+\sum_{j=1}^{K}I_j\Bigr),
  \label{eq:lin}
\end{equation}
the linearized recursion of Abou-Chacra, Anderson and
Thouless (AAT)~\cite{AAT1973}. The real recursion was studied in Ref.~\cite{parisi2019anderson}, where it was shown to contain information on the transition as well. Here we follow the more {\it traditional} approach and consider the flow of the imaginary part. Iterating it, the imaginary part at a site $o$ is a
sum over all the sites $x$ of the subtree below it,
\begin{equation}
  I_o=\eta\sum_x\;\prod_{k\in\mathcal{P}(o,x)}R_k^2
     =\eta\sum_x|G_{ox}|^2 ,
  \label{eq:I0sum}
\end{equation}
where $\mathcal{P}(o,x)$ is the unique path from $o$ to $x$ (both included)
and $G_{ox}=\prod_kR_k$ is the propagator from $o$ to $x$ at $\eta=0$. The
number of terms grows as $K^r$ with the depth $r$, while a typical term
decays exponentially in $r$. The sum is therefore dominated by rare paths with
anomalously large $|G_{ox}|$, and $I_o$ has a broad distribution with a power-law tail,
$P(I)\sim I^{-1-\mu}$. 

The localized solution, with $I\to0$ for $\eta\to0$, is stable as long as
this tail is thin enough. At finite $\eta$, however, the tail is a power law
only over a window, $\eta\lesssim I\lesssim1/\eta$, and its slope need not
be constant across it. We therefore define the local tail exponent through
the cumulative distribution $P_>(I)$ of $I$ at fixed $R$,
\begin{equation}
  m\equiv-\frac{d\ln P_>(I\,|\,R)}{d\ln I},
  \label{eq:mlocal}
\end{equation}
which tends to $\mu$ where the tail is a pure power law. The exponent
$m$ is the central quantity of this paper. We find its stationary values in Section~\ref{sec:lin-merger} and its flow in Section~\ref{sec:lin-flow}, where we also show that the $D_q$ depend on $\eta$ only through $m$, which is therefore the running coupling of the flow seen in Section~\ref{sec:popdyn}.

\subsection{The generating function}
\label{sec:lin-gen}

A heavy tail is most easily handled through the Laplace transform. Since
$R_i$ and $I_i$ are correlated, we define it at a given real part,
\begin{gather}
  \varphi(\tau\,|\,R)=\bigl\langle e^{-\tau I}\,\big|\,R\bigr\rangle \equiv \int_0^\infty dI\;e^{-\tau I}\,P(I\,|\,R),\nonumber \\ \chi(\tau\,|\,R)=1-\varphi(\tau\,|\,R).
\end{gather}
A tail $P(I)\simeq A\,I^{-1-\mu}$ with $0<\mu<1$ appears as a non-analytic
term at small $\tau$, $\chi\simeq A\,\Gamma(1-\mu)\,\tau^{\mu}/\mu$, while a
distribution with a finite mean gives an analytic $\chi\propto\tau$.
The same correspondence holds locally. With $s=\ln\tau$, if $m$ varies
slowly with $\ln I$, then $\chi(e^s\,|\,R)\simeq\Gamma(1-m)\,P_>(e^{-s}\,|\,R)$,
so that
\begin{equation}
  m(s)=\frac{\partial\ln\chi(e^s\,|\,R)}{\partial s}
  \label{eq:ms}
\end{equation}
is the local tail exponent of Eq.~\eqref{eq:mlocal} at $I\simeq e^{-s}$, up
to corrections of order $dm/ds$. The window $\eta\lesssim I\lesssim1/\eta$
becomes $|s|\lesssim\ell$, with $\ell=\ln(1/\eta)$.
The conditioning on $R$ is essential here. Fixing $R$ fixes how close the site itself is to resonance, and the remaining distribution of $I$ measures the broadening inherited from the subtree, which is what $m$ is meant to describe. Without conditioning, the average $\int dR\,P(R)\,\chi(\tau|R)$ contains a contribution from sites resonant with their own neighbors, which is $\propto(\eta\tau)^{1/2}$ at any disorder, and its local exponent would be $\frac12$ irrespective of $W$.

Given the real part $R_i$ at site $i$ and the real parts $R_j$ of its children, the $I_j$ are independent, since they come from disjoint subtrees. Eq.~\eqref{eq:lin} then gives
\begin{equation}
  \bigl\langle e^{-\tau I_i}\,\big|\,R_i,\{R_j\}\bigr\rangle
  =e^{-\tau R_i^2\eta}\prod_{j=1}^{K}\varphi\bigl(\tau R_i^2\,\big|\,R_j\bigr),
  \label{eq:phirec}
\end{equation}
where one generation multiplies the Laplace variable by $R_i^2$. Averaging over the $R_j$ and $V_i$ at fixed $R_i$ gives a closed equation for
$\varphi$, which is the linearized form of the self-consistency equation of
Refs.~\cite{Zirnbauer1986,MirlinFyodorov1991}. Near the localized solution
$\chi$ is small, and to first order
\begin{equation}
  \chi(\tau\,|\,R)=\tau R^2\eta
  +K\,\bigl\langle\chi(\tau R^2\,|\,R_1)\bigr\rangle_{R},
  \label{eq:chi}
\end{equation}
where $\langle\cdots\rangle_R$ averages over the child $R_1$, the other $K-1$ children and $V$, conditioned on the value $R$ of the parent. The factor $K$ in Eq.~\eqref{eq:chi} comes from the $K$ children being equivalent under the average $\langle \cdots \rangle_{R}$.

\subsection{Scaling modes and the merger of two fixed points}
\label{sec:lin-merger}

A tail $I^{-1-\mu}$ is a non-analytic term $\tau^\mu$ in $\chi$ (Section~\ref{sec:lin-gen}). We now ask which exponents Eq.~\eqref{eq:chi} allows. This is a standard analysis going back to \cite{AAT1973}, but we will revisit in slightly. 

The source $\tau R^2\eta$ is analytic in $\tau$: it fixes the $O(\tau)$ term (the mean of $I$ at given $R$) but cannot
generate a non-analytic term, which must therefore solve the homogeneous part
of Eq.~\eqref{eq:chi} by itself. That equation has no scale of its own:
$\tau$ enters only through the product $\tau R^2$, so one generation of the
tree acts on $\chi$ as a random rescaling of $\tau$, and if $\chi(\tau|R)$ is
a solution so is $\chi(\lambda\tau|R)$ for any $\lambda>0$. Its solutions can
therefore be classified by their behavior under dilatations of $\tau$, as power laws $\chi=\tau^{\mu}f_\mu(R)$. Substituting, the power of $\tau$
factors out and the profile $f_\mu$ must satisfy
\begin{equation}
  \bigl[\mathcal{T}(\mu)f_\mu\bigr](R)\equiv
  K|R|^{2\mu}\bigl\langle f_\mu(R_1)\bigr\rangle_R=\Lambda(\mu)\,f_\mu(R).
  \label{eq:Tmu}
\end{equation}
For each $\mu$, $\mathcal{T}(\mu)$ is a positive linear operator on functions of $R$. Since $\chi\geq0$, the profile $f_\mu$ must be positive, which selects the largest eigenvalue $\Lambda(\mu)$. The unknown is the exponent, which can be found by requiring that a tail with exponent $\mu$ is stationary, which is equivalent to enforcing $\Lambda(\mu)=1$ (the stationary values of the local exponent $m$ are the roots of $\Lambda(\mu)=1$).

The eigenvalue has a direct meaning in terms of the sum over paths,
Eq.~\eqref{eq:I0sum}:
\begin{equation}
  \Lambda(\mu)=\lim_{r\to\infty}\Bigl[\sum_{x\in\text{shell }r}
  \bigl\langle|G_{ox}|^{2\mu}\bigr\rangle\Bigr]^{1/r}.
  \label{eq:Lambda_paths}
\end{equation}
Going one generation deeper multiplies the number of paths by $K$ and their
weight $|G_{ox}|^{2\mu}$ by a factor $|R|^{2\mu}$, and $\Lambda(\mu)$ is the
net growth factor per generation of the $\mu$-th moment of the propagator.
The competition between the number of paths and their typical decay,
described below Eq.~\eqref{eq:I0sum}, is thus resolved moment by moment:
$\Lambda(\mu)<1$ means that even the rare paths with large $|G_{ox}|$ are too
few to make the $\mu$-th moment grow down the tree. In \cite{parisi2019anderson}, this was used to compute the susceptibility to a change in the local random energies and detect the transition.

The function $\Lambda(\mu)$ has been studied many times
before~\cite{AAT1973,MirlinFyodorov1991,KAI2018,parisi2019anderson}, and has
three important, but generic properties: First, $\Lambda(0)=K$, since $f=1$ is the
eigenvector at $\mu=0$; Second, $\ln\Lambda$ is convex in $\mu$; Third,
$\Lambda(\mu)=\Lambda(1-\mu)$, as shown by AAT~\cite{AAT1973}. This last point, we call the AAT-MF symmetry.

Hence $\Lambda(1)=K$, and the minimum of $\Lambda$ is at $\mu=\frac12$. Near the minimum, we can expand $\Lambda(\mu)\simeq\Lambda(\frac12)+\frac12\Lambda''\,(\mu-\frac12)^2$, and the roots of $\Lambda(\mu)=1$ are
\begin{equation}
  \mu_\pm=\frac12\pm\sqrt\delta,\qquad
  \delta\equiv\frac{2\bigl[1-\Lambda(\frac12)\bigr]}{\Lambda''(\frac12)} .
  \label{eq:mupm}
\end{equation}
Three possible cases can occur:
\begin{itemize}
  \item \emph{Localized}, $\Lambda(\frac12)<1$ ($\delta>0$): two real
        roots, paired by $\mu_++\mu_-=1$. The physical tail exponent is
        $\mu_+>\frac12$, as shown below.
  \item \emph{Critical}, $\Lambda(\frac12)=1$ ($\delta=0$): the two roots
        merge at $\mu=\frac12$. This is the mobility-edge criterion of
        Refs.~\cite{AAT1973,KAI2018}.
  \item \emph{Delocalized}, $\Lambda(\frac12)>1$ ($\delta=-\kappa^2<0$): the
        roots are complex, $\mu=\frac12\pm i\kappa$. The real solutions
        $\chi\propto\tau^{1/2}\cos(\kappa\ln\tau+\varphi_0)$ change sign,
        which a Laplace transform $1-\langle e^{-\tau I}\rangle$ cannot do: no
        stationary power-law tail exists, the linearization procedure we performed ceases to be valid and the nonlinear terms of Eq.~\eqref{eq:RI} take over.
\end{itemize}

From Eq.~\eqref{eq:I0sum} it is also possible to determine which of the two roots, $\mu_{+}$ and $\mu_{-}$, is realized in the localized phase. For $0<\mu\leq1$, $(\sum_xa_x)^\mu\leq\sum_xa_x^\mu$, so
\begin{equation}
  \bigl\langle I_o^{\mu}\bigr\rangle\leq\eta^{\mu}\sum_{r\geq0}\Lambda(\mu)^{r},
\end{equation}
which is finite whenever $\Lambda(\mu)<1$, i.e.\ for $\mu_-<\mu<\mu_+$. Lower
moments are then finite as well, since
$\langle I^{\mu'}\rangle\leq\langle I^{\mu}\rangle^{\mu'/\mu}$ for $\mu'<\mu$:
the bound fails below $\mu_-$, but the moments do not diverge there. All
moments of $I$ are therefore finite up to $\mu_+$, and the tail exponent is
$\mu_+$; the root $\mu_-$, with the heavier tail, is not reached from the
localized solution. In the language of Section~\ref{sec:lin-flow}, $\mu_+$ is
the infrared fixed point and $\mu_-$ its ultraviolet partner.
Although it is not reached, the root $\mu_-=1-\mu_+$ is observable: it sets how the localized phase is approached as $\eta\to0$. Rizzo, Tarzia and Vecchiarelli~\cite{RTV2026lett,RTV2026prb} have shown that the inverse participation ratio converges as $\mathrm{IPR}(\eta)-\mathrm{IPR}(0)\propto\eta^{\mu_+-\mu_-}=\eta^{2\mu_+-1}$, with an exponent fixed by the subdominant root, and that the correction becomes logarithmic in $\eta$ at the critical point, where the two roots merge. Since $\mu_+-\mu_-=2\sqrt\delta$, this is the same decay rate per unit $\ell=\ln(1/\eta)$ that we find for $D_1$ on the localized side, Section~\ref{sec:pd-results}.

Since $\Lambda(\frac12)$ is a smooth function of $W$, $\delta\propto W-W_c$, and the two exponents approach each other as $\sqrt{W-W_c}$. 

\subsection{The iteration equation as motion in AdS$_2$ and Conformality lost}
\label{sec:lin-ads}

As the disorder is varied, the exponents of the power-law solutions $\chi=\tau^\mu f_\mu(R)$ behave like the scaling dimensions of the operators dual to a scalar field in anti-de~Sitter (AdS) space: paired by $\mu\leftrightarrow1-\mu$, real in the localized phase, merging at $W_c$, complex in the delocalized phase. In AdS$_{d+1}$, a scalar of mass $M$ is dual to an operator of dimension $\Delta_\pm=\frac d2\pm\sqrt{d^2/4+M^2}$. The two dimensions are paired by $\Delta\leftrightarrow d-\Delta$. They are real above the Breitenlohner--Freedman (BF) bound $M^2_{\rm BF}=-d^2/4$, merge at the bound, and are complex below it. Kaplan et al.~\cite{Kaplan2009} identified the crossing of the BF bound as the holographic realization of conformality lost. There, $\Delta_-$ and $\Delta_+$ are the ultraviolet and infrared fixed points of a double-trace coupling of the boundary theory, and they merge at the bound. In a subsequent paper, Jensen et al.~\cite{JensenKarchSonThompson2010} found the holographic counterpart of the BKT transitions produced by the violation of the BF bound in an AdS$_2$ region of the geometry, with an exponentially small scale generated below the transition; see also Ref.~\cite{IqbalLiuMezei2015}. In this section we show that on the tree the analogy is more than qualitative: the linearized problem is a free scalar on AdS$_2$ ($d=1$).

Euclidean AdS$_2$ is the hyperbolic plane $\mathbb H_2$, the upper half-plane with the metric $ds^2=|dw|^2/(\operatorname{Im}w)^2$. The upper half-plane is where the propagators and the self-energies live: an iteration started with positive imaginary parts keeps them positive, and the population dynamics is the evolution of a distribution on it. The metric is natural as well. At $\eta=0$ and at fixed $\Sigma_i=\sum_jG_j$, one step of the recursion, $G_i=-1/(\Sigma_i-E+V_i)$, is a M\"obius map with unit determinant, and hence an isometry of $\mathbb H_2$. The AdS$_2$ structure that controls the tail, however, appears in a different half-plane, whose radial coordinate is the Laplace variable $\tau$; there, as is well known in AdS/CFT, power laws are the natural solutions.
 
In Poincar\'e coordinates $w=x+iz$, $z>0$, its metric and Laplacian are
\begin{equation}
  ds^2=\frac{dx^2+dz^2}{z^2},\qquad \nabla^2 \equiv \nabla^2_{\mathbb H_2} =z^2\bigl(\partial_x^2+\partial_z^2\bigr).
\end{equation}
Since the stationary modes factorize, $\chi=\tau^\mu f_\mu(R)$, Eq.~\eqref{eq:Tmu}, their dependence on $\tau$ can be studied by itself. We identify the radial coordinate $z$ with the Laplace variable $\tau$, i.e. $z=\tau=e^{s}$, and consider functions that do not depend on $x$. On such functions, $z\partial_z=\partial_s$, and $(z\partial_z)^2=z\partial_z+z^2\partial_z^2$, so that $z^2\partial_z^2=\partial_s^2-\partial_s$ and
\begin{equation}
  -\nabla^2=\partial_s\,(1-\partial_s),\qquad -\nabla^2\,\tau^{\mu}=\mu(1-\mu)\,\tau^{\mu} .
  \label{eq:lapH2}
\end{equation}

One step of the linearized cavity recursion multiplies a mode $\tau^\mu f_\mu(R)$ by $\Lambda(\mu)$, Eq.~\eqref{eq:Tmu}, and the mode is stationary if $\Lambda(\mu)=1$. By the AAT--MF symmetry, $\Lambda(\frac12+\varepsilon)$ is even in $\varepsilon$, so it depends on $\mu$ only through $\mu(1-\mu)=\frac14-\varepsilon^2$, and we define a new function $\tilde\Lambda$ as:
\begin{equation}
  \Lambda(\mu)=\tilde\Lambda\bigl(\mu(1-\mu)\bigr).
\end{equation}
Since $\mu(1-\mu)$ is the eigenvalue of $-\nabla^2$ on $\tau^\mu$, Eq.~\eqref{eq:lapH2}, the stationary modes are the $x$-independent solutions of
\begin{equation}
  \bigl[1-\tilde\Lambda(-\nabla^2)\bigr]\,\tau^{\mu}=0 .
  \label{eq:LH2}
\end{equation}
This follows from the reflection symmetry alone, which is exact; no expansion around $\mu=\frac12$ has been made. The operator contains all powers of $-\nabla^2-\frac14$, which equals $-(\mu-\frac12)^2$ on $\tau^\mu$; their coefficients are those of the expansion of $\Lambda$ in even powers of $\mu-\frac12$.

Near $\mu=\frac12$, $\Lambda\simeq\Lambda(\frac12)+\frac12\Lambda''\varepsilon^2$ with $\varepsilon^2=\frac14+\nabla^2$. Using the definition of $\delta$ in Eq.~\eqref{eq:mupm},
\begin{equation}
  1-\tilde\Lambda(-\nabla^2)\simeq\frac{\Lambda''}{2}\bigl[-\nabla^2+M^2\bigr],\qquad M^2=\delta-\frac14 ,
  \label{eq:KG}
\end{equation}
so that Eq.~\eqref{eq:LH2} becomes the Klein--Gordon equation of a free scalar of mass $M$ on AdS$_2$. Its $x$-independent solutions are $\tau^{\Delta}$. By Eq.~\eqref{eq:lapH2}, $\Delta(\Delta-1)=M^2$, which is the mass--dimension relation $\Delta(\Delta-d)=M^2$ at $d=1$. Its roots are $\Delta_\pm=\frac12\pm\sqrt{\frac14+M^2}$, and comparing with Eq.~\eqref{eq:mupm},
\begin{equation}
  \mu_\pm=\Delta_\pm ,\qquad M^2-M^2_{\rm BF}=\delta\propto W-W_c .
  \label{eq:AdSdict}
\end{equation}
The distance from the critical disorder plays the role of the squared mass measured from the bound, and the transition is the point where the bound is crossed. For $-\frac14<M^2<\frac34$ ($0<\delta<1$), both $\Delta_+$ and $\Delta_-$
are admissible and define two boundary theories, the infrared and the
ultraviolet one~\cite{Kaplan2009}; correspondingly, both $\mu_\pm$ lie in
$(0,1)$ and are legitimate tail exponents. Further roots of $\tilde\Lambda=1$, if any, and the non-Perron eigenvalues of $\mathcal T(\mu)$ correspond to heavier fields; they are irrelevant near the transition.

\begin{figure}[t]
\centering
\includegraphics[width=\columnwidth]{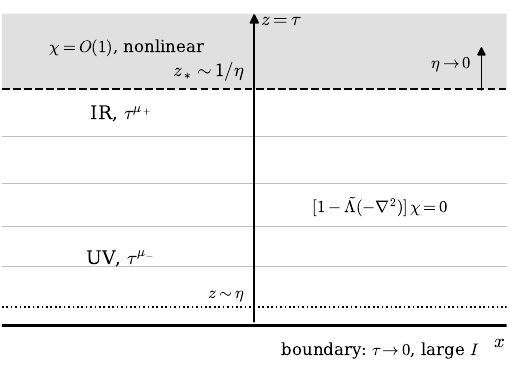}
\caption{The Poincar\'e half-plane of Euclidean AdS$_2$, with the radial coordinate $z$ identified with the Laplace variable $\tau$. The generating function $\chi$ does not depend on $x$. Near the boundary, $\tau\to0$ probes the far tail of $P(I)$ and the fall-off $\tau^{\mu_-}$ (ultraviolet) dominates. Towards the bulk, the typical values of $I$ and $\tau^{\mu_+}$ (infrared) dominate. The linear equation holds in the window $\eta\lesssim z\lesssim1/\eta$, where $\chi$ obeys Eq.~\eqref{eq:LH2}. Beyond the wall at $z_*\sim1/\eta$ (dashed), $\chi=O(1)$ and the nonlinear terms enter. As $\eta\to0$ the wall moves into the interior, at unit rate in $\ln z$ (Section~\ref{sec:lin-flow}).}
\label{fig:ads2}
\end{figure}

The identification $z=\tau$ fixes the rest of the dictionary. The boundary $z\to0$ is $\tau\to0$, which probes the far tail of $P(I)$, at large $I$. The interior $z\to\infty$ is large $\tau$, which probes the typical, small values of $I$. The two fall-offs $z^{\Delta_-}$ and $z^{\Delta_+}$ will turn out to be the ultraviolet and infrared fixed points of the flow of Section~\ref{sec:lin-flow}, as in holography. The geometry is sketched in Fig.~\ref{fig:ads2}.

The boundary coordinate $x$ plays no role, since $\chi$ does not depend on it. The $x$-independent solution $\tau^\mu$ is the bulk-to-boundary propagator $K_\mu(z,x;x_0)=[z/(z^2+(x-x_0)^2)]^\mu$ from the boundary point $x_0=\infty$: indeed, $\lim_{x_0\to\infty}x_0^{2\mu}K_\mu=z^\mu$.

The same structure appears in the supersymmetric formulation~\cite{Zirnbauer1986,MirlinFyodorov1991}. There the order parameter is a function on $\mathbb H_2$ that is radial about the point singled out by $\eta$. The eigenfunctions of the linearized self-consistency operator are the cone functions $\mathcal P_{-1/2+ip}$, with eigenvalue $\Lambda(\frac12+ip)$, and $\mu\leftrightarrow1-\mu$ is the identity $\mathcal P_\sigma=\mathcal P_{-1-\sigma}$. Far from that point, the geodesic distance is $\rho\simeq-s$ up to a constant, and radial functions approach the $x$-independent ones used here.

In summary, the homogeneous part of the linearized cavity recursion, Eq.~\eqref{eq:chi}, is invariant under dilatations of $\tau$, so its solutions are power laws $\chi = \tau^\mu f_\mu(R)$, which one cavity step multiplies by $\Lambda(\mu)$, Eq.~\eqref{eq:Tmu}. Stationarity requires $\Lambda(\mu)=1$, and the AAT--MF symmetry makes $\Lambda$ a function $\tilde\Lambda$ of $\mu(1-\mu)$, which is also the eigenvalue of $-\nabla^2$ on $\tau^\mu$. Consequently, the stationarity condition can be written exactly as $[1-\tilde\Lambda(-\nabla^2)]\,\chi=0$, an equation involving only the Laplacian of $\mathbb{H}_2$, with $\tau$ as the radial coordinate. Near $\mu=\frac12$ it reduces to the Klein--Gordon equation of a free scalar on AdS$_2$ whose squared mass, measured from the BF bound, is $\delta\propto W-W_c$: the localized phase lies above the bound, the transition at it, and the delocalized phase below it, where conformality is lost. 
At finite $\eta$, $\chi$ is a superposition of the two fall-offs, and its running exponent is the boundary condition on a cutoff surface at the wall (Section~\ref{sec:lin-flow}). The correspondence does not involve a gravitational bulk, and there is no dynamics in $x$. How much of the conformal group $\mathrm{SL}(2,\mathbb{R})$ acts on $\chi$, and how the merger at $W_c$ becomes a Jordan block of the dilatation, is discussed in Appendix~\ref{app:conformal}.

\subsection{From the eigenvalue to a flow}
\label{sec:lin-flow}

The exponents found above are $\eta$-independent and describe the limit $\eta\to0$. The RG/finite-size flow describes how this limit is approached, through the dependence of the running exponent $m$ of Eq.~\eqref{eq:mlocal} on the scale. At finite $\eta$, $\chi$ is small and non-analytic only in the window $\eta\lesssim\tau\lesssim1/\eta$, i.e.\ $|s|\lesssim\ell$ with $\ell=\ln(1/\eta)\simeq\ln N+\text{const}$ (Section~\ref{sec:lin-gen}), which lies below the wall of Fig.~\ref{fig:ads2} and widens as the system grows.

Multiplying $\tau$ by $R^2$ translates $s$ by $\ln R^2$, so the homogeneous
part of Eq.~\eqref{eq:chi} acts on $\chi$ as a random translation in $s$.
Formally, with $\partial_s$ the generator of $s$-translations,
$\chi(\tau R^2\,|\,R_1)=e^{\ln R^2\,\partial_s}\chi(s\,|\,R_1)
=|R|^{2\partial_s}\chi(s\,|\,R_1)$, and the homogeneous equation reads
\begin{equation}
  \begin{gathered}
  \chi=\mathcal{T}(\partial_s)\chi ,\\
  \bigl[\mathcal{T}(\partial_s)\chi\bigr](s,R)
  =K\bigl\langle\chi(s+\ln R^2\,|\,R_1)\bigr\rangle_R ,
  \end{gathered}
\end{equation}
which is the transfer operator~\eqref{eq:Tmu} with $\mu$ replaced by
$\partial_s$. On a single power law, $\partial_s$ acts as multiplication by
$\mu$: the modes $e^{\mu s}f_\mu(R)$ are eigenfunctions of
$\mathcal{T}(\partial_s)$ with eigenvalue $\Lambda(\mu)$, and the stationary
solutions of Section~\ref{sec:lin-merger} are recovered. At finite $\eta$,
however, $\chi$ need not be a single power law across the window; it is a
superposition of such modes, and its local slope is the running exponent
$m(s)$.

Near the transition only modes with $\mu$ close to $\frac12$ matter, since
$\mu_\pm-\frac12=\pm\sqrt\delta$ is small. The eigenvector $f_\mu(R)$ is
smooth in $\mu$, so all these modes share the same profile in $R$ up to
corrections of order $\sqrt{|\delta|}$, and we look for
\begin{equation}
  \chi(s,R)\simeq f_{1/2}(R)\,h(s) .
  \label{eq:factor}
\end{equation}
This factorization holds only in a region of the $(R,s)$ plane: the
translation by $\ln R^2$ must keep $s$ inside the window, $R^2\eta\tau\ll1$,
which holds uniformly in the window only for $|R|<R_c$ with $R_c=O(1)$. Since the typical $|R|$ is $\sim1/W$, with a Cauchy tail
$P(R)\simeq1/(WR^2)$ beyond it, the cut excludes only a fraction
$\simeq2/(WR_c)\approx0.1$ of the sites already for $R_c=1$. These few sites, however, dominate the full average: since $f_\mu(R)\sim|R|^{2\mu}$, $\langle\chi(\tau\,|\,R)\rangle_R$ is controlled for $\mu\ge\frac12$ by sites resonant with their neighbors, $|R|\sim(\eta\tau)^{-1/2}$, which give a term $\propto(\eta\tau)^{1/2}$ whatever $\delta$. The profile $h$ is therefore read from the conditional generating function
\begin{equation}
  \begin{aligned}
  \bar\chi(s)&\equiv\chi\bigl(e^{s}\,\big|\,|R|<R_c\bigr)\\
  &=1-\bigl\langle e^{-e^{s}I}\,\big|\,|R|<R_c\bigr\rangle\propto h(s).
  \end{aligned}
  \label{eq:chibar}
\end{equation}

Acting with $\mathcal{T}(\partial_s)$ on Eq.~\eqref{eq:factor}, each
Fourier--Laplace component $e^{\mu s}$ of $h$ is multiplied by $\Lambda(\mu)$,
and the $R$-dependence drops out, leaving $[1-\Lambda(\partial_s)]\,h=0$. With $\Lambda(\mu)=\tilde\Lambda(\mu(1-\mu))$ and $-\nabla^2=\partial_s(1-\partial_s)$, Eq.~\eqref{eq:lapH2}, this is Eq.~\eqref{eq:LH2} for an arbitrary superposition of modes, $[1-\tilde\Lambda(-\nabla^2)]\,h=0$. To second order around $\mu=\frac12$, Eq.~\eqref{eq:KG} gives $-\nabla^2+M^2=\delta-(\partial_s-\frac12)^2$, so
\begin{equation}
  \Bigl(\partial_s-\frac12\Bigr)^2h=\delta\,h,
  \qquad
  h(s)=A\,e^{\mu_+s}+B\,e^{\mu_-s}.
  \label{eq:hode}
\end{equation}

The two roots of $\Lambda(\mu)=1$ have become the two independent solutions
of a second-order equation, the Klein--Gordon equation of Section~\ref{sec:lin-ads} written in $s=\ln z$. With $h=e^{s/2}\psi$ it reads $\psi''=\delta\,\psi$.

The higher-order terms, proportional to $(\partial_s-\frac12)^4$, shift
$\mu_\pm$ by a relative $O(\delta)$. Replacing a translation by $\ln R^2$
with derivatives is justified deep in the window, where $h$ varies on the
scale $|\delta|^{-1/2}$, much longer than the typical jump $\ln R^2$. The
coefficients $A$ and $B$ are not fixed by Eq.~\eqref{eq:hode}; we return to them below. Since both are generically nonzero, $h$ crosses over
from one power law to the other as $s$ varies across the window: since $\mu_-<\mu_+$, $B\,e^{\mu_-s}$ dominates towards the AdS boundary ($s\to-\infty$, large $I$) and $A\,e^{\mu_+s}$ towards the interior ($s\to+\infty$, typical $I$). In the
region where Eq.~\eqref{eq:factor} holds, the local exponent of
Eq.~\eqref{eq:ms} no longer depends on $R$, $m(s)=d\ln h/ds$, and this
crossover makes it run.

\begin{figure*}[t]
\centering
\includegraphics[width=\textwidth]{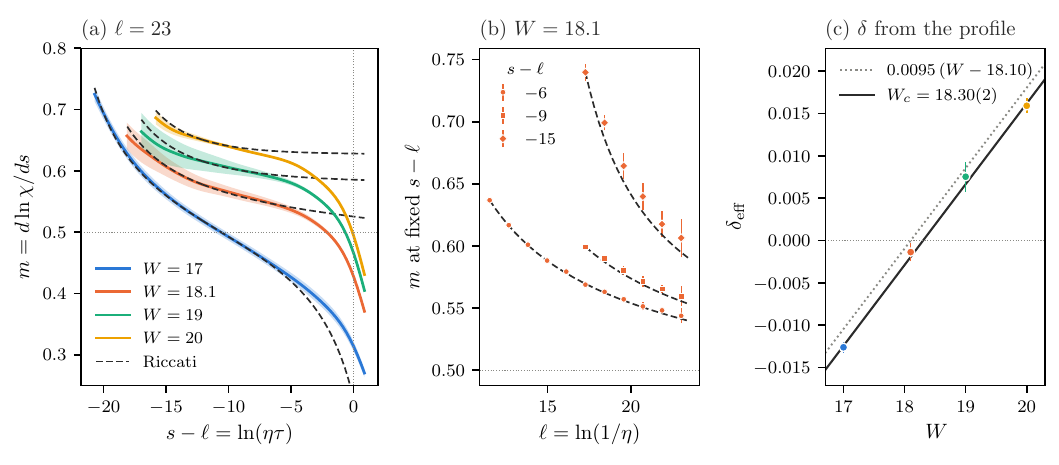}
\caption{Local exponent $m=d\ln\bar\chi/ds$ of the conditional generating
function~\eqref{eq:chibar}, from the population dynamics ($K=2$, $E=0$,
$N=10^6$, $R_c=1$) compared with the solution of the Riccati equation \eqref{eq:riccati}.
(a)~As a function of $s-\ell=\ln(\eta\tau)$ at $\ell=23$ for $W=17$, $18.1$,
$19$, $20$; bands are two standard errors. (b)~At $W=18.1$, at fixed
$s-\ell$ as a function of $\ell=\ln(1/\eta)$. Dashed: Eq.~\eqref{eq:riccati},
in $s$ in~(a) and in $\ell$ in~(b), started from a data point, with
$\delta=a\,(W-W_c)$ from~(c). (c)~$\delta$ from the residual of
Eq.~\eqref{eq:riccati} for $-16<s-\ell<-8$; solid: linear fit; dotted: the
line of Eq.~\eqref{eq:pd-fit}.}
\label{fig:chiR}
\end{figure*}

Since $m-\frac12=d\ln\psi/ds$, $\psi''=\delta\psi$ gives $d(m-\frac12)/ds=\psi''/\psi-(\psi'/\psi)^2$, which is the Riccati equation
\begin{equation}
  \frac{dm}{ds}=\delta-\Bigl(m-\frac12\Bigr)^2=-(m-\mu_+)(m-\mu_-),
  \label{eq:riccati}
\end{equation}
which is the normal form of Ref.~\cite{Kaplan2009} with coupling $m$. As $s$ grows, $m$ flows to
$\mu_+$, the infrared fixed point, away from $\mu_-$, the ultraviolet one.

This flow has a direct holographic meaning. On the surface $z=\tau$ of AdS$_2$, $m-\frac12=d\ln\psi/ds$ is the ratio of the radial derivative of the bulk field to its value, which is the coefficient of a Robin boundary condition. In AdS/CFT, the boundary condition relating the coefficients of $z^{\Delta_-}$ and $z^{\Delta_+}$ is a double-trace coupling of the boundary theory~\cite{Witten2001multitrace}. When the cutoff surface is moved into the bulk, this coupling obeys a Riccati equation, which follows from the radial Hamilton--Jacobi equation~\cite{HeemskerkPolchinski2011,FaulknerLiuRangamani2011}. Equation~\eqref{eq:riccati} is this flow, with $m-\frac12$ as the double-trace coupling and $\mu_\mp=\Delta_\mp$ as its two fixed points, the two quantizations of the bulk field~\cite{Kaplan2009}.

The solutions are
\begin{equation}
  \begin{aligned}
  m&=\frac12+\sqrt\delta\, \coth \bigl[\sqrt\delta(s-s_0)\bigr],\\
  m&=\frac12+\frac{1}{s-s_0},\\
  m&=\frac12-\kappa\tan\bigl[\kappa(s-s_0)\bigr],
  \end{aligned}
\end{equation}
in the localized, critical, and delocalized cases, respectively. In the localized case $m$ approaches $\mu_+$ from above, as in Fig.~\ref{fig:chiR}(b); the other solution, $\frac12+\sqrt\delta\tanh[\sqrt\delta(s-s_0)]$, lies between $\mu_-$ and $\mu_+$ and would give $D_1<0$. In the delocalized solution, $m$ runs from $+\infty$ to $-\infty$ over $\Delta s=\pi/\kappa$, crossing $\frac12$ halfway: the linear regime lasts a scale range $\propto(W_c-W)^{-1/2}$, and then the nonlinear terms take over.
This is the walking regime of Ref.~\cite{Kaplan2009}, and the origin of $\ln N_\xi\propto(W_c-W)^{-1/2}$ for the correlation volume $N_\xi$~\cite{Zirnbauer1986,MirlinFyodorov1991}. Over this range of scales, the system looks nonergodic, which accounts for the stretched-exponential decay of the return probability and the subdiffusive spreading observed at intermediate times on the delocalized side~\cite{bera2018return,detomasi2020subdiffusion}.

As $\eta$ decreases, the window grows, and observables measured at its upper end read the profile at a moving point. In Figure~\ref{fig:chiR} we show the local exponent of $\bar\chi$, Eq.~\eqref{eq:chibar}, measured with the population dynamics of Section~\ref{sec:popdyn} ($N=10^6$, $R_c=1$) at four values of $W$. The exponent $m(\ell)$ obeys Eq.~\eqref{eq:riccati} with $s$ replaced by $\ell=\ln(1/\eta)$, provided this point moves through the profile at unit rate. A scaling argument fixes the rate. Away from the upper end of the window, Eq.~\eqref{eq:chi} is invariant under translations of $s$, and $\eta$ enters only through $\eta\tau=e^{s-\ell}$: in the source term, $\tau R^2\eta=R^2e^{s-\ell}$, and in the nonlinear region at the wall. There the relevant variables are $I/\eta$ and $b/\eta$, and for $\eta\to0$ the full recursion for them does not depend on $\eta$, except through resonances $|E-V_i-\sum_jR_j|\lesssim b$, whose weight vanishes with $\eta$. The wall therefore sits at a fixed value of $s-\ell$: observables read $m$ at $s_*=\ell+\sigma$, with $\sigma$ independent of $\eta$, and $ds_*/d\ell=1$, as shown in  Fig.~\ref{fig:chiR}(b). In the critical solution $h\propto(A+Bs)\,e^{s/2}$ and $s_0=-A/B$, so the ratio of the two components is fixed where the profile is anchored, at an $\eta$-independent scale near the beginning of the flow, and $\ell_0=\sigma-s_0$ is of order one, as in the data of Section~\ref{sec:popdyn}. The wall fixes instead the normalization $\chi_+$ of the profile, which sets the coefficient $c$ of the flow of $D_1$ below. In the language of Section~\ref{sec:lin-ads}, the wall is a cutoff surface at $z_*=e^{\sigma}/\eta$ in the interior of AdS$_2$. Lowering $\eta$ moves it inwards at unit rate in $\ln z$, and since $\eta\sim1/N$, its radial position is $\ln z_*=\ln N+\text{const}$.

Away from the wall, $s-\ell\ll0$, $m(s)$ follows Eq.~\eqref{eq:riccati}: at $W=17$ it crosses $\frac12$ and keeps decreasing, at $W=19$ and $20$ it decreases towards $\mu_+$, and at $W=18.1$ it decreases slowly towards $\frac12$. For each $W$ we take $\delta$ from the residual $dm/ds+(m-\frac12)^2$, averaged over $-16<s-\ell<-8$ at $\ell=23$.
It is linear in $W$, $\delta=a\,(W-W_c)$, with $a=0.0095(2)$ and $W_c=18.30(2)$; an independent run on a narrower grid gives $a=0.0098(2)$ and $W_c=18.25(2)$. The slope agrees with that of Eq.~\eqref{eq:pd-fit}, obtained from the flow of $D_1$. The two runs together give $W_c=18.27(3)$, above the value $W_c\simeq18.1$ of the fit of $D_1$ by many standard errors. We attribute the difference to the finite value of $\eta$, through the cut $R_c$. Increasing $R_c$ will move $W_c$ estimated this way closer to its real value but would also require considerable more computational effort. 

The conditional profile therefore confirms the Riccati form of the flow and the slope $a$. Near the wall, $s-\ell\gtrsim-5$, the profile bends down: there $\chi=O(1)$ and the linearization no longer holds. Deeper than $s-\ell\simeq-20$, the measurement is limited by $N$, since a population of $N$ sites cannot represent $\chi\lesssim10/N$.

\begin{figure*}[ht]
\centering
\includegraphics[width=0.9\textwidth]{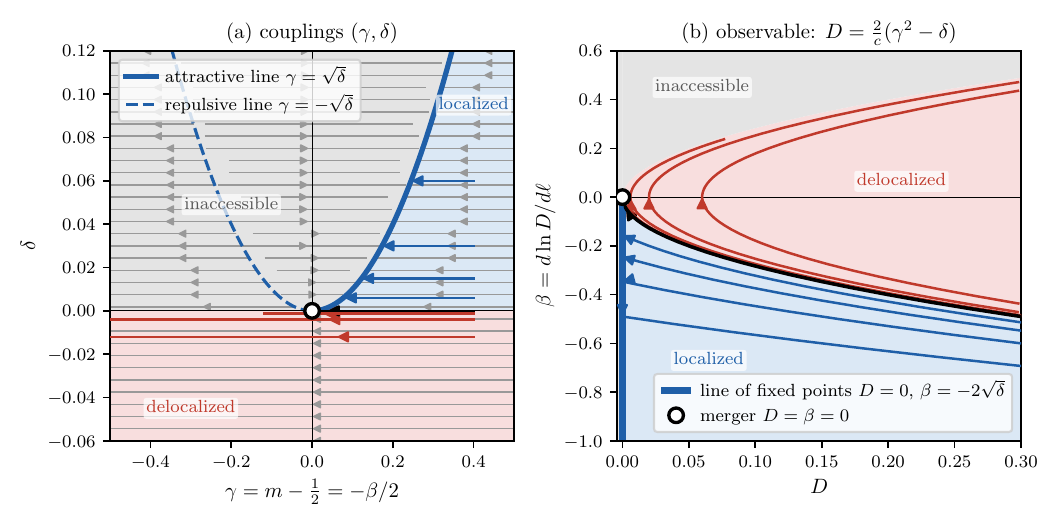}
\caption{Topology of the flow on the tree. Trajectories start at
$\gamma=0.4$ (blue: localized; black: critical, $\delta=0$; red:
delocalized). (a)~The plane of couplings $(\gamma,\delta)$, with streamlines in
grey: the attractive line $\gamma=\sqrt\delta$ and the repulsive line $\gamma=-\sqrt\delta$ merge at
$\delta=0$. (b)~The same trajectories in the plane of the observable,
$(D,\beta=d\ln D/d\ell)$, with $D$ from Eq.~\eqref{eq:D1beta}, shown only
for $D\le0.3$, close to the origin, where the linear analysis holds. The tree has a
line of fixed points, $D=0$, $\beta=-2\sqrt\delta$, which is the line
$\gamma=\sqrt\delta$ of~(a), ending at the merger $D=\beta=0$. The value $c=0.4$ is
illustrative. Shading: localized (light blue), delocalized (light red) and
inaccessible (grey) regions; trajectories started at large $\gamma$ are stopped by the
attractive line and never enter the grey region.}
\label{fig:treeflow}
\end{figure*}

\paragraph{Observables.} The fractal dimensions which enter the observables are read at the upper end of the window where the linear approximation holds. After integrating out the site energy, $D_1=d\langle\ln(b/\eta)\rangle/d\ell$, with $b$ the imaginary part of the self-energy, and $\langle\ln(b/\eta)\rangle$ is a functional of the profile $\bar\chi$ below the wall. By Eq.~\eqref{eq:hode} this profile depends on $\ell$ only through $m(\ell)$, and the functional turns out to be exactly linear in $m(\ell)$ (Appendix~\ref{app:obs-tree}). With Eq.~\eqref{eq:riccati} this gives
\begin{equation}
  \begin{aligned}
  D_1(\ell)&=-\frac2c\,\frac{dm(\ell)}{d\ell}=\frac2c\Bigl[\Bigl(m(\ell)-\frac12\Bigr)^2-\delta\Bigr],\\
  \beta(\ell)&=\frac{d\ln D_1(\ell)}{d\ell}=1-2m(\ell) ,
  \end{aligned}
  \label{eq:D1beta}
\end{equation}
the relations used in Section~\ref{sec:popdyn}, and Eq.~\eqref{eq:riccati} becomes the two-parameter flow of Ref.~\cite{Vanoni2024},
\begin{equation}
  \frac{d\ln D_1}{d\ell}=\beta,\qquad \frac{d\beta}{d\ell}=c\,D_1,\qquad
  \frac{\beta^2}{2}-c\,D_1=2\delta .
  \label{eq:betaflow}
\end{equation}
The same flow, with the same invariant and its continuation to $\delta<0$, was obtained independently and simultaneously in Refs.~\cite{RTV2026lett,RTV2026prb}, from the finite-size scaling of the inverse participation ratio. Eq.~\eqref{eq:D1beta} clarifies the meaning of the two variables of Ref.~\cite{Vanoni2024}: $\beta$ is the running coupling itself, measured from the self-dual point $m=\frac12$, and $D_1$ is its beta function, up to the factor $-2/c$. The fixed points of $m$ are therefore the zeros of $D_1$: the infrared one at $\beta=-2\sqrt\delta$, and the ultraviolet one at $\beta=+2\sqrt\delta$. On the localized line of fixed points $\beta=-(\mu_+-\mu_-)\propto(W-W_c)^{1/2}$, which replaces the value $-1/(\xi\ln K)$ proposed in Ref.~\cite{Vanoni2024}. For $\delta<0$ the parabola detaches from the axis, $D_1\ge2|\delta|/c$, and the flow walks. The coefficient $c$, Eq.~\eqref{eq:cdef}, is set at the wall by $\chi_+$ and is a parameter of the flow. Ref.~\cite{RTV2026prb} writes the same flow for $D_2$, with coefficient $g=2I_2^\infty/\Phi_0$, where $\Phi_0$ is the amplitude of the $1/\ln N$ correction to the inverse participation ratio; since $D_2=r_2D_1$, $c=r_2\,g$. With their $g\simeq1.2$ at $W_c$ and our $r_2\simeq0.35$ this gives $c\simeq0.42$, consistent with our fits, $c\simeq0.4$ (Section~\ref{sec:popdyn}). For the Anderson model $\Phi_0$ is fitted to the data in Ref.~\cite{RTV2026prb}, so this relates $c$ to another non-universal amplitude rather than predicting it. Every $D_q$ with $q>\frac12$ follows the same flow with the same $\beta$, $D_q=r_qD_1[1+O(m-\mu_+)]$ (Appendix~\ref{app:obs-tree}).

The flow equations of Ref.~\cite{Vanoni2024} thus follow directly from the self-consistency equations of the Bethe lattice, through the dependence of their solution on the regulator $\eta$, interpreted as $\eta\propto1/N$~\cite{RTV2026lett}.

\section{The flow at large $d$: a $1/d$ expansion}
\label{sec:largeD}

The analysis of this section is of a different nature from that of Section~\ref{sec:linear}. There, the flow was derived from the cavity equations of the Bethe lattice. Here, we do not derive the $1/d$ corrections from a microscopic expansion: we write the minimal deformation of the tree flow allowed by dimensional counting and by the two effects of the loops, with two coefficients, $a$ and $b$, which we do not compute and which we fix by comparison with the numerical exponents. The equations below are therefore phenomenological, and are meant as a starting point for a $1/d$ program.

In Refs.~\cite{Vanoni2024,Altshuler2025}, the fractal dimensions played the role of a coupling. This was an {\it escamotage} to find the topology of the RG flow. However, as discussed above, $D_q$ are observables and not couplings. They are functions of the couplings, but the RG equations should be written in terms of the couplings themselves. The couplings are now identified as the running exponent $m$ and the rate $\delta$. These are the only slowly varying modes of the theory on the tree: $\delta$ is immobile, fixed by $W$, and $m$ (equivalently, $h$) is the only slow mode, since the real part of the propagator converges quickly to its distribution $P(R)$ and, after decoupling, the operator $\mathcal{T}$ is gapped.

It is convenient to measure the running exponent from the self-dual point
$\mu=\frac12$ of the AAT symmetry $\Lambda(\mu)=\Lambda(1-\mu)$, and we define
\begin{equation*}
  \gamma\equiv m-\frac12 ,
\end{equation*}
so that $\beta=1-2m=-2\gamma$ by Eq.~\eqref{eq:D1beta}, and the fixed points $m=\mu_\pm$ are $\gamma=\pm\sqrt\delta$. On the tree, the RG equations are
\begin{equation}
  \frac{d\gamma}{d\ell}=\delta-\gamma^2,\qquad
  \frac{d\delta}{d\ell}=0 .
\end{equation}
The topology of this flow is shown in Fig.~\ref{fig:treeflow}: the blue curves correspond to the localized phase and the red curves to the delocalized phase.

The rate $\delta$ is a constant, and this is a peculiarity of the tree. How can we deform the flow to get the flow in finite but large dimensions? Here we need to specify the rules for counting engineering dimensions around the $d\to\infty$ limit.

With $\epsilon=1/d$, the equation for $\delta$ needs to change. At the very least, one has to consider its bare, or {\it engineering}, dimension. To obtain it, note that on the tree, changing the volume $N$ by a factor $N\to e^{u}N$, as in the RG transformation we have defined, changes its diameter by a constant, $L\to L+u/\ln K$, so the scaling dimension of a length is zero.

Now, on the localized line of fixed points $D\propto e^{-2\sqrt\delta\,\ell}$, so $2\sqrt\delta$ is a decay rate per unit $\ell$. On a regular lattice, the scaling dimension of a length is 1, so the scaling dimension of $\delta$ is 2. Defining the flow as $d/d\ln N$, which allows the limit $d\to\infty$ to be taken smoothly, the engineering dimension of $\delta$ must be $2/d=2\epsilon$. This reasoning must be done for every coupling of the RG. Fortunately, there are only two, $\gamma$ and $\delta$, and $\gamma$, being a running exponent, is dimensionless for any $d$.

Keeping the definition of $\ell=\ln N$, we have a minimal deformation of the flow:
\begin{equation}
  \frac{d\gamma}{d\ell}=\delta-\gamma^2,\qquad
  \frac{d\delta}{d\ell}=2\epsilon\,\delta .
  \label{eq:eps-flow}
\end{equation}
The only fixed point is now $(\gamma,\delta)=(0,0)$, where the Jacobian of the flow is
\begin{equation}
  J=\begin{pmatrix}
    \partial_\gamma\dot\gamma & \partial_\delta\dot\gamma\\
    \partial_\gamma\dot\delta & \partial_\delta\dot\delta
  \end{pmatrix}_{(0,0)}
  =\begin{pmatrix}0&1\\0&2\epsilon\end{pmatrix},
  \label{eq:eps-jacobian}
\end{equation}
with the dot denoting $d/d\ell$. It is triangular, so its eigenvalues are the diagonal
entries, $0$ and $2\epsilon$. The right eigenvectors give the directions in the
plane of couplings: $(1,0)$, with eigenvalue $0$, is the critical line
$\delta=0$, and $(1,2\epsilon)$, with eigenvalue $2\epsilon$, is the direction
$\delta=2\epsilon\gamma$, tangent to the unstable manifold discussed below. The
left eigenvectors give instead the scaling fields, the combinations of the
couplings that the linearized flow multiplies by $e^{\lambda\ell}$: these are
$u_t=\delta$, with eigenvalue $2\epsilon$, and $u_\gamma=\gamma-\delta/(2\epsilon)$, with
eigenvalue $0$. Indeed $du_\gamma/d\ell=(\delta-\gamma^2)-\delta=-\gamma^2$, which
vanishes at linear order.
Along $\delta$ the eigenvalue is $2\epsilon$, or 2 in units of $\ln L$
(since $\ell=d\ln L$): $\delta$ is relevant and $\nu=\frac12$. Along $\gamma$ it is
0: $\gamma$ is marginal. Beyond linear order, on the critical line
$d\gamma/d\ell=-\gamma^2$: for $\gamma>0$ it is marginally irrelevant and flows to
zero as on the tree, $\gamma=1/(\ell+\ell_0)$, which gives logarithmic
corrections to scaling; for $\gamma<0$ it runs away.

\paragraph{Adding the loops.} The flow~\eqref{eq:eps-flow} has $\nu=\frac12$ and a
marginal $\gamma$, while in finite $d$ the numerics give $\nu>\frac12$ and an
irrelevant direction. On a lattice the loops can act in two different ways.
At a fixed scale they break the reflection symmetry $\Lambda(\mu)=\Lambda(1-\mu)$
of the tree: $G_{ox}$ is now a sum over many interfering paths, and $\Lambda$
might acquire a term odd in $\mu-\frac12$, which enters the equation for $\gamma$ as
a term linear in $\gamma$. A loop can also {\it decrease} the rate of decay of the wave function, since it can increase the probability of a resonance. This adds a term $\propto-\delta$ to the second equation. Both terms must be subleading in $\epsilon$ with respect to the engineering dimension, and the simplest flow\footnote{One can see that adding a term $\epsilon\gamma\delta$ to $d\gamma/d\ell$ does not change anything.} containing both effects is
\begin{equation}
  \frac{d\gamma}{d\ell}=\delta-\gamma^2-a\epsilon\,\gamma,\qquad
  \frac{d\delta}{d\ell}=\bigl(2\epsilon-b\epsilon^2\bigr)\,\delta ,
  \label{eq:eps-flow-loops}
\end{equation}
with $a$ and $b$ two numbers of order one. The equation for $\delta$ still
forces $\delta^*=0$ at a fixed point, and then $\gamma^*(\gamma^*+a\epsilon)=0$: there
are two fixed points, $(\gamma,\delta)=(0,0)$ and $(-a\epsilon,0)$. The merger of the
tree has been split and we have recovered conformality. The Jacobian at the critical point $(0,0)$ is
\begin{equation}
  J=\begin{pmatrix}-a\epsilon&1\\0&2\epsilon-b\epsilon^2\end{pmatrix},
  \label{eq:eps-jacobian-loops}
\end{equation}
still triangular, because $d\delta/d\ell$ does not depend on $\gamma$, so its
eigenvalues are again the diagonal entries,
\begin{equation}
  \lambda_t=2\epsilon-b\epsilon^2,\qquad \lambda_\gamma=-a\epsilon .
\end{equation}
The corresponding right eigenvectors are $(1,0)$ for $\lambda_\gamma$, the critical
line $\delta=0$ as before, and $\bigl(1,\epsilon(2+a-b\epsilon)\bigr)$ for $\lambda_t$, so
that the unstable manifold now leaves the fixed point along
$\delta\simeq\epsilon(2+a)\gamma$. The scaling fields are $u_t=\delta$ and
$u_\gamma=\gamma-\delta/[\epsilon(2+a-b\epsilon)]$.

\begin{figure*}[t]
\centering
\includegraphics[width=0.9\textwidth]{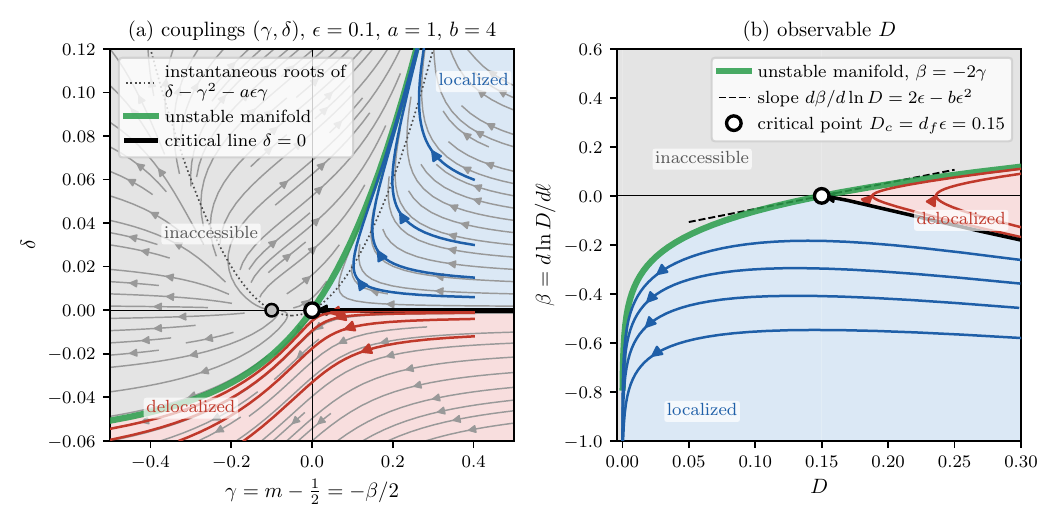}
\caption{The flow with the loop terms, Eq.~\eqref{eq:eps-flow-loops}, with
$\epsilon=0.1$, $a=1$, $b=4$; same initial conditions and colors as in
Fig.~\ref{fig:treeflow}. Shading: localized (light blue), delocalized (light
red) and inaccessible (grey) regions; the unstable manifold is invariant, so
trajectories started at large $\gamma$ stay on its right and never enter the grey
region. (a)~The plane of couplings: the merger is split into
the critical point $(0,0)$ (white) and its ultraviolet partner $(-a\epsilon,0)$
(grey), and the critical line now flows into $(0,0)$ along an irrelevant
direction, with eigenvalue $-a\epsilon$. The unstable manifold (green) is
$\gamma=\psi'/\psi-a\epsilon/2$ with $\psi=I_n(\zeta)$ for $\delta>0$ and $J_n(\zeta)$ for
$\delta<0$, $\zeta=\sqrt{|\delta|}/\epsilon'$, $\epsilon'=\epsilon-b\epsilon^2/2$,
$n=a\epsilon/(2\epsilon')$. (b)~The plane of the observable, for $D\le0.3$, with
$D=T+\epsilon d_f\Phi$ as in Eq.~\eqref{eq:eps-D} and
$\Phi=[(\zeta/2)^n/(\Gamma(n+1)I_n(\zeta))]^2$, which reduces to Eq.~\eqref{eq:eps-Phi}
for $a=b=0$ and makes $\beta=-2\gamma$ exact on the manifold; the critical point
is at $D_c=d_f\epsilon$, and the dashed tangent
at $D_c$ has slope $d\beta/d\ln D=2\epsilon-b\epsilon^2$, giving $\nu=1/(2-b/d)$. The
coefficients $c=0.4$ and $d_f=1.5$ are illustrative.}
\label{fig:epsflowloops}
\end{figure*}

Multiplying by $d$ to convert to units of $\ln L$, the relevant eigenvalue is
\begin{equation}
  \begin{gathered}
  y_t=d\,\lambda_t=2-\frac bd,\\
  \nu=\frac1{y_t}=\frac{1}{2-b/d}=\frac12+\frac{b}{4d}+O(d^{-2}),
  \end{gathered}
  \label{eq:eps-nu-loops}
\end{equation}
so a positive $b$ gives $\nu>\frac12$, with a correction of order $1/d$. The value $b=4$ gives exactly $\nu=1/(2-4/d)=\frac12+1/(d-2)$, the form proposed in Ref.~\cite{Altshuler2025}, $1/\nu=2(1-2/d)$. This value is suggested both by the numerics and by the conjecture $D_c=2/d$. The numerical exponents in $d=3$--$6$, $\nu=1.57$, $1.11$, $0.96$, $0.84$~\cite{Tarquini17}, are close to the values $1.50$, $1.00$, $0.83$, $0.75$ given by $b=4$, and the linear fit $1/\nu\simeq2-4.75/d$ of Ref.~\cite{Tarquini17} corresponds to $b\simeq4.75$; and with $D_c=2/d$ ($d_f=2$), the same relation reads $1/\nu=2(1-D_c)$. The other eigenvalue is
\begin{equation}
  y=d\,\lambda_\gamma=-a ,
  \label{eq:eps-y-loops}
\end{equation}
so for $a>0$ the coupling $\gamma$ is irrelevant: the logarithmic corrections to scaling of the minimal flow become powers, $L^{y}=L^{-a}$. The same numerics find $y\simeq-1$ in $d=3$ and $4$, and $-1.2$, $-1.4$ in $d=5$, $6$~\cite{Tarquini17}, which suggests $a\simeq1$. We actually conjecture that $a=1, d_f=2$ exactly. The two numbers play separate roles. $b$ changes only $\nu$ and $a$ changes only $y$; in particular, the breaking of the reflection symmetry alone, however strong, cannot change $\nu$, because it does not touch the equation for $\delta$. The second fixed point, $(-a\epsilon,0)$, has eigenvalues $+a\epsilon$ and $\lambda_t$: it is fully repulsive, the ultraviolet partner of the critical point, at a distance $a\epsilon$ from it. On the tree, $\epsilon=0$, the two coincide, and we are back to the merger. The resulting flow is shown in Fig.~\ref{fig:epsflowloops}.

Besides the critical line, a second invariant line departs from the fixed point
along the relevant direction, $\delta\simeq\epsilon(2+a-b\epsilon)\gamma$. It is the one-parameter scaling curve, the green line in Fig.~\ref{fig:epsflowloops}, and it has a closed form. The linear term in $\gamma$ is removed by the shift $\gamma=\psi'/\psi-a\epsilon/2$: since $\gamma'=\psi''/\psi-(\psi'/\psi)^2$ and $\delta-\gamma^2-a\epsilon\gamma=\delta-(\psi'/\psi)^2+a^2\epsilon^2/4$, Eq.~\eqref{eq:eps-flow-loops} becomes
\begin{equation}
\label{eq:BesselEq}
  \psi''=\Bigl(\delta+\frac{a^2\epsilon^2}{4}\Bigr)\psi,\qquad
  \delta\propto e^{2\epsilon'\ell},\quad \epsilon'=\epsilon-\frac{b\epsilon^2}{2}.
\end{equation}
In the variable $\zeta=\sqrt{|\delta|}/\epsilon'$, for which $d/d\ell=\epsilon' \zeta\,d/d\zeta$, Eq.~\eqref{eq:BesselEq} is Bessel's equation of order $n=a\epsilon/(2\epsilon')$: modified for $\delta>0$, ordinary for $\delta<0$. This $\psi$ is the same function as in Section~\ref{sec:lin-flow}: for $\epsilon\to0$, with $\ell$ in place of $s$, Eq.~\eqref{eq:BesselEq} reduces to $\psi''=\delta\,\psi$, and $\gamma=\psi'/\psi=m-\frac12$, since $h=e^{s/2}\psi$ gives $m=d\ln h/ds=\frac12+\psi'/\psi$.
The invariant line is the solution that departs from the critical point, the one regular at $\zeta=0$: $\psi=I_n(\zeta)\propto \zeta^n$ for $\delta>0$, and
$\psi=J_n(\zeta)\propto \zeta^n$ for $\delta<0$. This gives
\begin{equation}
  \gamma=
  \begin{cases}
    \displaystyle\epsilon'\,\frac{\zeta\,I_n'(\zeta)}{I_n(\zeta)}-\frac{a\epsilon}{2}, & \delta>0,\\[10pt]
    \displaystyle\epsilon'\,\frac{\zeta\,J_n'(\zeta)}{J_n(\zeta)}-\frac{a\epsilon}{2}, & \delta<0.
  \end{cases}
  \label{eq:eps-manifold}
\end{equation}
At $\zeta\to0$, $\zeta I_n'/I_n\to n$ and $\gamma\to\epsilon'n-a\epsilon/2=0$, as it
should. The second, independent solution of Eq.~\eqref{eq:BesselEq} for
$\delta>0$, $\psi=K_n(\zeta)\propto \zeta^{-n}$, gives instead $\zeta K_n'/K_n\to-n$ and
$\gamma\to-a\epsilon$. Since $\zeta\to0$ corresponds to $\ell\to-\infty$, it
describes a line that departs from the ultraviolet fixed point
$(-a\epsilon,0)$. Every combination $\psi=A\,I_n+B\,K_n$ with $B\neq0$ does
the same, so the unstable manifold of the critical point, $B=0$, is the
only trajectory that does not come from it. 

The two branches of Eq.~\eqref{eq:eps-manifold} are one analytic function of $\delta$, since $I_n(iy)=i^nJ_n(y)$ and $\zeta\,d/d\zeta$ is unchanged by $\zeta\to i\zeta$. For
$\sqrt\delta\gg\epsilon$, $\zeta I_n'/I_n\simeq \zeta-\frac12$ and $\gamma\simeq\sqrt\delta-(\epsilon'+a\epsilon)/2$: the line approaches the attractive root $-a\epsilon/2+\sqrt{\delta+a^2\epsilon^2/4}\simeq\sqrt\delta-a\epsilon/2$, trailing it by $\epsilon'/2$. For $\delta<0$ it has a pole at the first zero of $J_n$ ($\zeta\simeq3.32$ for the values of Fig.~\ref{fig:epsflowloops}), discussed in Appendix~\ref{app:obs-1d}. Since $\gamma$ is now irrelevant, generic trajectories approach this line as a power of $L$, with the exponent $y=-a$, and not as $1/\ln\xi$. For $a=b=0$ we have $n=0$ and $\epsilon'=\epsilon$, and Eq.~\eqref{eq:eps-manifold} reduces to the line of the minimal flow~\eqref{eq:eps-flow}, $\gamma=\sqrt\delta\,I_1(\zeta)/I_0(\zeta)$ for $\delta>0$ and
$\gamma=-\sqrt{|\delta|}\,J_1(\zeta)/J_0(\zeta)$ for $\delta<0$, with $\zeta=\sqrt{|\delta|}/\epsilon$
and the pole at $\zeta=2.405$.

\paragraph{Observables.} The fractal dimensions are not couplings but functions on the plane of couplings, carried along by the flow (Appendix~\ref{app:obs-1d}). On the tree $D$ is, up to a factor, the beta function of $\gamma$, Eq.~\eqref{eq:D1beta}, and vanishes at any fixed point: $D_c=0$. On a lattice, a critical support that is a fractal of dimension $d_f$ in real space contains $L^{d_f}=N^{d_f/d}$ sites, and the measured $D$ acquires a finite critical value, while the exponent is read from the slope at that point:
\begin{equation}
  D_c=\frac{d_f}{d},\qquad
  \frac1\nu=d\,\frac{d\beta}{d\ln D}\bigg|_{D_c}=2-\frac bd ,
  \label{eq:eps-slope}
\end{equation}
in agreement with Eq.~\eqref{eq:eps-nu-loops}. $D_c$ belongs to the observable, $\nu$ to the flow, as for the critical conductance in the scaling theory of Ref.~\cite{AALR1979}. Deep in the localized phase the growing $\delta$ gives $\beta\simeq\frac1d\ln D$, the signature of exponential localization; on the delocalized side the linearized flow sends $\gamma\to-\infty$ at $L\sim\xi$, where the linearization fails and the trajectory turns onto the ergodic arc (Appendix~\ref{app:obs-1d}).

\begin{figure}[!t]
\centering
\includegraphics[width=\columnwidth]{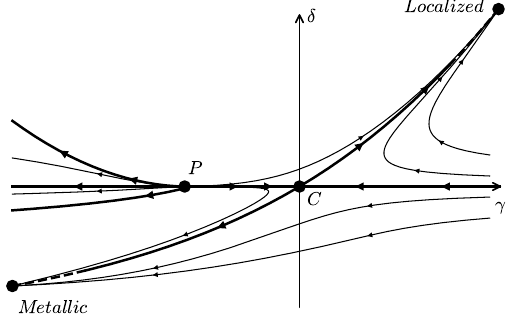}
\caption{Schematic depiction of the RG flow, conjectured at any finite $d$ in the plane of couplings
$(\gamma,\delta)$, $\gamma=m-\frac12$. The critical point $C$ is attractive along $\gamma$
and repulsive along its unstable manifold, which runs to the localized phase
and to the metallic (ergodic, $D=1$) phase, both at infinity on this scale (dashed ends and dots). $P$:
the ultraviolet partner of $C$ (at $\gamma=-a\epsilon$ for large $d$), repulsive in both
directions, with invariant lines. Thin lines: generic trajectories, leaving $P$ or entering from
large $\gamma$. At $d=\infty$, $P$ and $C$ merge.}
\label{fig:schematic}
\end{figure}

\section{Conclusions}
\label{sec:conclusions}
We have shown that in the limit of $d\to\infty$, the Anderson transition loses its CFT character and is described by a BKT-like flow, which is a special case of the {\it conformality lost} scenario of Kaplan et al.~\cite{Kaplan2009}, which appeared in other contexts in high-energy physics. We also find that the AdS/CFT picture of Ref.~\cite{Kaplan2009} is naturally realized in the Anderson model: the order parameter of the supersymmetric formulation lives on the hyperbolic plane $\mathbb{H}_2$, and the slow mode of Zirnbauer, and of Mirlin and Fyodorov, is a scalar field in $\mathrm{AdS}_2$. The Anderson transition occurs where the mass of this scalar field crosses the BF bound. This implies that the two solutions of the stationary condition $\Lambda(\mu)=1$ are two fixed points of the RG flow, which has two couplings: $\delta$ (the decay rate) and the running exponent $m$ (the local power law of the distribution of the local density of states). We have used this as a starting point for an $\epsilon=1/d$ expansion of the critical theory of the Anderson transition, showing that the rate has dimension $2$ and the exponent $m$ is marginal, and have put forward a simple picture for the modification of the RG equations at $O(\epsilon)$ which accounts for the numerical values of the critical exponent $\nu$ and of the first irrelevant exponent $y$ for $d\gtrsim 4$. We have identified two terms in the RG equations, which are the effects of loops on the exponent of the distribution of the local density of states, and the increase in the number of resonances, which reduces the rate of decay of the wave function with the distance.

The resulting picture is summarized in Fig.~\ref{fig:schematic}. At $d=\infty$, the two fixed points of $m$ merge at $m=\frac12$ and annihilate as $W$ decreases through $W_c$: this is the conformality-lost scenario. At finite $d$ the loops separate them: the critical point $C$ stays at $\gamma=0$, where $\gamma$ is now irrelevant, with eigenvalue $-a\epsilon$, while the disorder $\delta$ is relevant,
with eigenvalue $2\epsilon-b\epsilon^2$, so that $1/\nu=2-b/d$, Eq.~\eqref{eq:eps-nu-loops}; its ultraviolet partner $P$ sits at $\gamma=-a\epsilon$
and is repulsive in both directions. The unstable manifold of $C$ connects the
critical point to the two phases. On the localized side it runs to
$\gamma,\delta\to\infty$. On the delocalized side, the linearized flow
\eqref{eq:eps-flow-loops} sends $\gamma\to-\infty$ at a finite scale, where the
linearization around the localized solution no longer holds; the full flow runs instead to the metallic phase.

It is a task for future work to derive the RG equations from the field theory, possibly the nonlinear $\sigma$-model of Wegner and Efetov~\cite{Wegner1979,Efetov1983}, and to fix the values of their coefficients (or to add more terms to the equations).

Related results were obtained independently and simultaneously by Rizzo, Tarzia and Vecchiarelli~\cite{RTV2026lett,RTV2026prb}, who derive the BKT flow and its coefficients analytically from the finite-size scaling of the inverse participation ratio, in agreement with their numerics, and show that the finite-$N$ and finite-$\eta$ corrections share the same exponent.

\begin{acknowledgments}
We would like to thank B. L. Altshuler and V. E. Kravtsov for collaborations and discussions that inspired the present work. We also thank Federico Balducci, Soumya Bera, and Yan Fyodorov for comments on the first version of this paper. The code used to obtain the results reported in this work has been generated with the assistance of Claude and run on computational resources managed and supported by Princeton Research Computing. Claude has also been used for literature search and for substantially refining the theoretical picture. We have independently verified the arguments we presented, and we take full responsibility for their validity.
\end{acknowledgments}

\appendix

\section{Observables}
\label{app:observables}

This appendix collects the derivations of the results on the observables summarized in Sections~\ref{sec:lin-flow} and~\ref{sec:largeD}.

\subsection{Observables on the tree}
\label{app:obs-tree}

All the fractal dimensions $D_q$ with $q>\frac12$, like $D_1$, are determined by the exponent at the upper end of the window. In fact, after integrating out the site energy, as in Section~\ref{sec:pd-method}, $\ln I=\ln\cos^2\theta-\ln b$, where $b=\eta+\sum_{j=1}^{K+1}I_j$ is the imaginary part of the self-energy. The first term does not depend on $b$, and for $b\ll W$ neither do the weights $(\theta_2-\theta_1)/W$. Using the Shannon-entropy definition, Eq.~\eqref{eq:pd-D1} then gives
\begin{equation}
  D_1=\frac{d}{d\ell}\bigl\langle\ln(b/\eta)\bigr\rangle.
  \label{eq:D1b}
\end{equation}
Therefore, $D_1$ is the rate at which the logarithm of the {\it typical value} of the self-energy moves with $\eta$.

We can now connect $D_1$ to the analytic solution for $m(\ell)$, and write down the equations in Ref.~\cite{Vanoni2024} for the flow near the origin of $(D, \beta)$. Since $b=\eta+\sum_{j=1}^{K+1}I_j$ with independent $I_j$, the identity
$\ln(b/\eta)=\int_0^\infty(d\tau/\tau)\,(e^{-\tau\eta}-e^{-\tau b})$ gives
\begin{equation}
  \bigl\langle\ln(b/\eta)\bigr\rangle\simeq\int_{-\infty}^{\infty}ds\;e^{-\eta e^{s}}
  \Bigl[1-\bigl(1-\bar\chi(s)\bigr)^{K+1}\Bigr],
  \label{eq:lnbexact}
\end{equation}
with $\bar\chi$ given by Eq.~\eqref{eq:chibar}. With $\bar\chi$ replaced by the average of $\chi(e^s\,|\,R)$ over all $R$,
Eq.~\eqref{eq:lnbexact} would be exact. Using $\bar\chi$ is nevertheless a
good approximation: a child with $|R_j|>R_c$ only multiplies its
$I_j=R_j^2\,b_j$ by a large factor, whose distribution does not depend on
$\eta$. To leading order this shifts $\langle\ln(b/\eta)\rangle$ by an
$\eta$-independent constant, which drops out of Eq.~\eqref{eq:D1b}.
Here $e^{-\eta e^{s}}$ acts as the wall at $s\simeq\ell$. 
In particular, the self-resonant term $\propto(\eta\tau)^{1/2}$ of the unconditioned average contributes $\int^{\ell}ds\,e^{(s-\ell)/2}=O(1)$.

At this point, to estimate the fractal dimension $D_1$, we take the wall as a step at $s=\ell$, keep the window profile $\bar\chi(s)=\chi_+\,h(s)/h(\ell)$ for $-\ell<s<\ell$, with $\chi_+=O(1)$ being the value of $\bar\chi$ at the wall, set $\bar\chi=0$ outside, and work to first order in $\bar\chi$. Because $h$ solves the second-order equation~\eqref{eq:hode}, its shape below the wall is fixed by $\delta$ and by $m(\ell)$ alone:
\begin{equation}
  \begin{aligned}
  \frac{h(\ell+u)}{h(\ell)}={}&e^{u/2}\Bigl[\cosh\bigl(\sqrt\delta\,u\bigr)\\
  &+\frac{m(\ell)-\frac12}{\sqrt\delta}\,\sinh\bigl(\sqrt\delta\,u\bigr)\Bigr],
  \qquad u\le0 .
  \end{aligned}
  \label{eq:hratio}
\end{equation}
For $\mu_->0$, the integrand decays away from the wall, the lower end of the window contributes only $O(e^{-2\mu_-\ell})$, and Eq.~\eqref{eq:lnbexact} reduces to
\begin{equation}
  \begin{aligned}
  \bigl\langle\ln(b/\eta)\bigr\rangle(\ell)&\simeq(K+1)\,\chi_+\int_{-\infty}^{0}du\,
  \frac{h(\ell+u)}{h(\ell)}\\
  &=(K+1)\,\chi_+\,\frac{1-m(\ell)}{\mu_+\mu_-},\\
  &\qquad\text{with}\quad \mu_+\mu_-=\tfrac14-\delta .
  \end{aligned}
  \label{eq:lnbm}
\end{equation}
At the fixed point $m(\ell)=\mu_+$ this reduces to $(K+1)\chi_+/\mu_+$, as for a
pure power law. Eq.~\eqref{eq:lnbm} is exactly linear in $m(\ell)$: writing
$1-m(\ell)=(1-\mu_+)-[m(\ell)-\mu_+]$ and using $1-\mu_+=\mu_-$, we get
\begin{equation}
  \bigl\langle\ln(b/\eta)\bigr\rangle(\ell)=\frac{(K+1)\,\chi_+}{\mu_+}
  -\frac2c\,\bigl[m(\ell)-\mu_+\bigr],
  \label{eq:lnblin}
\end{equation}
where
\begin{equation}
  c=\frac{2\mu_+\mu_-}{(K+1)\,\chi_+}
  \;\xrightarrow{\;W=W_c\;}\;\frac{1}{2(K+1)\,\chi_+} .
  \label{eq:cdef}
\end{equation}
Consequently, $c$ is the response of the mean self-energy to the running exponent; it is set by the value $\chi_+$ of the profile at the wall, and $c>0$ because a thinner tail means fewer broadened sites. The step is not essential: since $h(\ell+u)/h(\ell)$ is linear in $m(\ell)$, a smooth wall of $\eta$-independent width in $s-\ell$ leaves Eq.~\eqref{eq:lnblin} linear in $m(\ell)$ and only renormalizes $\chi_+$, hence $c$.
Eqs.~\eqref{eq:riccati}, \eqref{eq:D1b} and~\eqref{eq:cdef} give Eqs.~\eqref{eq:D1beta} and~\eqref{eq:betaflow} of Section~\ref{sec:lin-flow}.

The same argument applies to every $D_q$ with $q>\frac12$. Integrating out
the site energy gives $\langle I^q\rangle\propto\langle b^{1-q}\rangle$, so
that
\begin{equation}
  D_q=\frac{1}{1-q}\,\frac{d}{d\ell}\ln\bigl\langle(b/\eta)^{1-q}\bigr\rangle ,
  \label{eq:Dqb}
\end{equation}
which reduces to Eq.~\eqref{eq:D1b} for $q\to1$. Writing $(b/\eta)^{1-q}=\Gamma(q-1)^{-1}\int_0^\infty(d\tau/\tau)\,(\eta\tau)^{q-1}e^{-\tau b}$ for $q>1$ (and a similar representation with $1-e^{-\tau b}$ for $\frac12<q<1$), as for the logarithm, makes the average a functional of the profile $\bar\chi$ below the wall, which by Eq.~\eqref{eq:hratio} depends
on $\ell$ only through $m(\ell)$. Hence $D_q\propto dm/d\ell$ for every $q$,
$D_q=r_q\,D_1[1+O(m-\mu_+)]$, and all $D_q$ follow the flow~\eqref{eq:betaflow}
with the same $\beta=1-2m$.

\subsection{Observables in $1/d$}
\label{app:obs-1d}

\paragraph{The localized phase.} In the localized phase, we need to recover an exponential decay for observables such as the $D_q$. As explained in Ref.~\cite{Altshuler2025}, this requires $\beta=\frac{1}{d}\ln D_q+O(1)$. We recover this now for our deformed flow.

On the localized side $\delta$ is positive and, by Eq.~\eqref{eq:eps-flow-loops}, it is no longer stationary: it grows as $  \delta(\ell)=\delta_0\,e^{2\epsilon'\ell}$, with $\epsilon'=\epsilon-b\epsilon^2/2$. Since $\ell=d\ln L$, we have $e^{2\epsilon'\ell}=L^{y_t}$ with
$y_t=2-b/d=1/\nu$, so $\sqrt\delta=\sqrt{\delta_0}\,L^{1/(2\nu)}$ grows with the size of the system. The coupling $\gamma$ cannot stay behind: for $\delta>0$ its equation has the attractive root $\gamma=-a\epsilon/2+\sqrt{\delta+a^2\epsilon^2/4}\simeq\sqrt\delta-a\epsilon/2$, towards which $\gamma$ relaxes at the rate $\simeq2\sqrt\delta$. Once $\sqrt\delta\gg\epsilon$, the root moves slowly on this scale, and $\gamma$ simply tracks it, lagging behind by $\epsilon'/2$ (see Eq.~\eqref{eq:eps-manifold}); these shifts of order $\epsilon$ can be neglected here, and $\gamma\simeq\sqrt\delta$. Through Eq.~\eqref{eq:D1beta} this fixes the beta function, $\beta=-2\gamma\simeq-2\sqrt{\delta_0}\,e^{\epsilon'\ell}$, and since
$\beta=d\ln D/d\ell$ we can integrate it to find $D$ as a function of $\delta$:
\begin{equation}
  \begin{aligned}
  \ln D&=\int\beta\,d\ell\simeq-2\sqrt{\delta_0}\int e^{\epsilon'\ell}\,d\ell\\
  &=-\frac{2\sqrt{\delta_0}}{\epsilon'}\,e^{\epsilon'\ell}+\text{const}
  =-\frac{2\sqrt\delta}{\epsilon'}+\text{const}.
  \end{aligned}
\end{equation}
Comparing the two expressions, $\beta\simeq-2\sqrt\delta$ and
$\ln D\simeq-2\sqrt\delta/\epsilon'$, we can eliminate $\delta$ and obtain, up to an
additive constant, $\beta\simeq\epsilon'\ln D$. For $b=0$, $\epsilon'=\epsilon$ and this is
\begin{equation}
\frac{d\ln D}{d\ell}=\frac{1}{d}\ln D,
\end{equation}
whose solution is $D=e^{-(L-L_0)/\xi}$ with $\xi$ arbitrary, which is just what one
expects from an exponentially localized state.

For $b\neq0$ there is a subtlety. In terms of $L$, the same result reads $\ln D\simeq-2(L/\xi)^{1/(2\nu)}$, with $\xi=(\epsilon'/\sqrt{\delta_0})^{2\nu}\propto
(W-W_c)^{-\nu}$: for $\nu>\frac12$ this is a stretched exponential, not the
exponential decay of a localized state. The reason is that
Eq.~\eqref{eq:eps-flow-loops} keeps the anomalous dimension of $\delta$ all the way
into the localized phase. The anomalous dimension is a property of the critical
point. Deep in the localized phase, $\delta\gg\epsilon^2$, $\sqrt\delta$ is a decay rate
per unit distance and must scale exactly as a length, with dimension $2$, so
the term $b\epsilon^2\delta$ must switch off once $L\gtrsim\xi$. Beyond this crossover
$\sqrt\delta\simeq\epsilon L/\xi$, and the argument above gives back $D\propto e^{-2L/\xi}$ and
$\beta=\frac1d\ln D$, now with $\xi\propto(W-W_c)^{-\nu}$. The flow with a constant
$b$, used in Fig.~\ref{fig:epsflowloops}, is accurate near the critical point,
which is where it is needed for $\nu$. The crossover can be built into the flow by letting the anomalous term depend on $\delta$,
\begin{equation}
  \frac{d\delta}{d\ell}=2\epsilon\,\delta-b\epsilon^2\,\delta\,g\bigl(\delta/\epsilon^2\bigr),\qquad g(0)=1,\quad g(x)\xrightarrow{x\to\infty}0,
  \label{eq:eps-flow-g}
\end{equation}
for instance $g(x)=1/(1+x)$. Near the critical point $g\simeq1$ and Eq.~\eqref{eq:eps-flow-loops} is recovered with the same eigenvalues, since only the linearization at $\delta=0$ enters $\nu$; deep in the localized phase, $\delta\gg\epsilon^2$, the flow returns to $d\delta/d\ell\simeq2\epsilon\delta$ and the exponential decay of $D$ is restored.

It is instructive to compare this with what happens on the tree. There
$\epsilon=0$, $\delta$ does not move, and $\beta=-2\sqrt\delta$ stays constant along
each trajectory, so that $D\propto e^{-2\sqrt\delta\,\ell}=N^{-2\sqrt\delta}$. Since on
the tree the distance is itself proportional to $\ell$, this is also an
exponential decay with the distance, but it is a power of $N$, and each value
of $W$ comes to rest at its own fixed point $\beta=-2\sqrt\delta$: this is the
line of fixed points of Fig.~\ref{fig:treeflow}. On the lattice, instead, the
distance is $L=e^{\epsilon\ell}$, $\beta=\epsilon\ln D$ keeps decreasing, and every
localized trajectory runs off to $D\to0$, $\beta\to-\infty$. The line of fixed
points of the tree has been replaced by a single attractor at $\beta=-\infty$.
\paragraph{The delocalized phase.} For $\delta<0$ the picture is different.
Now $\delta=-|\delta|$ stays negative while $|\delta|$ grows as
$e^{2\epsilon'\ell}$. At fixed $\delta$, the flow of $\gamma$,
$d\gamma/d\ell=-(\gamma^2+a\epsilon\gamma+|\delta|)$, has two fixed points,
$\gamma=\frac12\bigl[-a\epsilon\pm\sqrt{a^2\epsilon^2-4|\delta|}\bigr]$, both
negative: they are the critical point and its ultraviolet partner, displaced
by $\delta$. As $|\delta|$ grows, they approach each other, and at
$|\delta|=a^2\epsilon^2/4$ they merge at $\gamma=-a\epsilon/2$ and disappear.
This is the merger of the tree, which the loops have moved from $\delta=0$ to
a finite $|\delta|$ that the flow itself reaches. Beyond it, the right-hand
side is negative for every $\gamma$, and $\gamma$ decreases without bound,
reaching $-\infty$ at a finite scale. Along the invariant line,
Eq.~\eqref{eq:eps-manifold}, this happens at the first zero of $J_n$, where
$\psi\to0$: $\zeta=\sqrt{|\delta|}/\epsilon'\simeq3.32$ for the values of
Fig.~\ref{fig:epsflowloops} ($2.405$ for $a=b=0$), that is, at a finite size
with $L/\xi$ of order one. Correspondingly $\beta=-2\gamma\to+\infty$ and $D$ grows without bound, which signals that the linearized flow has broken down.

This pole is a fiction: it comes from extending the linear approximation far
from the critical point, where it no longer holds. Eq.~\eqref{eq:eps-flow-loops} is
the flow linearized around the localized solution, to first order in
the imaginary part $\chi$. As $\gamma$ decreases the tail of $P(I)$ becomes
heavier, $\chi$ is no longer small, and well before the pole $D$ reaches
values of order one, where the linearization breaks down; $D$, of course,
can never exceed one. The same happens on the tree, where in the delocalized
case $m=\frac12-\kappa\tan[\kappa(s-s_0)]$ runs to $-\infty$ in a finite range and
the nonlinear terms of Eq.~\eqref{eq:RI} take over (Section~\ref{sec:lin-flow}).
What the pole does tell us correctly is the scale at which the flow leaves
the critical region, $L\sim\xi$. Beyond it the trajectory turns onto the
ergodic arc and approaches $D=1$, as in Section~\ref{sec:popdyn}
($\beta\simeq\lambda(1-D_1)$ with $\lambda\simeq1$ on the tree); this part of the flow
is nonlinear and is not described by Eq.~\eqref{eq:eps-flow-loops}.

\paragraph{From the RG flow to the observables.}
In Fig.~\ref{fig:epsflowloops}(b), the critical point sits at a finite $D_c=d_f\epsilon$, while in the plane of couplings the fixed point is at $(\gamma,\delta)=(0,0)$, exactly as on the tree. The two statements are compatible because they concern different objects. The flow acts on the couplings, and a fixed point is a point $(\gamma^*,\delta^*)$ in their plane. $D$ is not a coupling: it is a measured quantity, a function $\mathcal D(\gamma,\delta)$ on that plane, which the flow carries along. Its value at the fixed point, $D_c=\mathcal D(\gamma^*,\delta^*)$, depends on which function one measures; the exponents, which are eigenvalues of the linearized flow, do not. This is familiar. At the Anderson transition $D_1$ and $D_2$ take different values at the same fixed point, and on the tree $D_q/D_1=r_q\neq1$ at every $\ell$: the $D_q$ are different functions on one plane of couplings. Likewise, in the scaling theory of Ref.~\cite{AALR1979} the critical conductance depends on how the conductance is defined, while $\nu$ does not.

On the tree, the definition forces $D_c=0$. By Eq.~\eqref{eq:D1beta},
\begin{equation}
  D=\frac2c\bigl(\gamma^2-\delta\bigr)=-\frac2c\,\frac{d\gamma}{d\ell}:
\end{equation}
$D$ measures the speed at which the coupling $\gamma$ runs: up to a factor, it is the beta function of $\gamma$, and this vanishes at any fixed point. No
deformation of the flow alone can change this. On the tree $D_c=0$ is
correct: at $W_c$, $D=2/[c(\ell+\ell_0)^2]\to0$.

The lattice changes the relation between $D$ and $L$, and this has to be taken into account {\it on top} of the modification of the RG equations for the couplings.
Going back to the counting definition of $D$,
$D=d\ln\mathcal N/d\ln N$, with $\mathcal N$ the number of sites on which the
eigenfunction lives: we have $\mathcal N\propto N^D$. Suppose that at
the critical point this support is a fractal in real space of dimension
$d_f$~\cite{Altshuler2025}, so that a box of side $L$ contains $\mathcal N\propto L^{d_f}$ of its
sites. On a lattice $N=L^d$, hence
\begin{equation}
  \begin{aligned}
  \mathcal N&\propto L^{d_f}=N^{d_f/d},\\
  \ln\mathcal N&=\frac{d_f}{d}\,\ln N+\text{const},\qquad
  D=\frac{d_f}{d} .
  \end{aligned}
\end{equation}
A power of the distance is now a power of $N$ (unlike on the tree), and the fractal support gives a
finite $D$. In general $\mathcal N$ also contains a factor $F$ controlled by the couplings, $\mathcal N=L^{d_f} F$, and $\ln\mathcal N$ is the sum of the two logarithms, so $D=d_f/d+d\ln
F/d\ln N$. The support stops growing once $L$ exceeds $\xi$, and the
observable takes the form
\begin{equation}
  \mathcal D(\gamma,\delta)=T(\gamma,\delta)+\epsilon d_f\,\Phi(\delta),
  \label{eq:eps-D}
\end{equation}
and the measured value at scale $\ell$ is
$D(\ell)=\mathcal D\bigl(\gamma(\ell),\delta(\ell)\bigr)$. The second term
is the fractal support: $\Phi$ depends on the scale only through $\delta(\ell)$, via $\zeta=\sqrt{|\delta|}/\epsilon'$, with $\Phi=1$ at
the critical point and, in the localized phase, $\Phi\to0$ for $L\gg\xi$
($\zeta\gg1$).

The first term measures the transient: how far $\gamma$ is from the
one-parameter curve, and must vanish on it. On the tree this role is played
by $D$ itself, $D=(2/c)(\gamma-\sqrt\delta)(\gamma+\sqrt\delta)$: the product
of the distances from the two fixed lines $\gamma=\pm\sqrt\delta$, which on
the tree are exact solutions because $\delta$ does not run. With the loops
$\delta$ runs, and the fixed lines are no longer solutions: $\gamma$ trails
the attractive one, $\simeq\sqrt\delta-a\epsilon/2$, by $\epsilon'/2$, and
the tree formula would read this lag as a speed and give a negative $D$ in
the localized phase. We therefore measure the distances from the two exact
solutions of the linearized flow instead: the invariant line $\gamma_I$ of
Eq.~\eqref{eq:eps-manifold}, and the line $\gamma_K$ obtained with $K_n$ in
place of $I_n$, which departs from the ultraviolet fixed point. For
$\delta>0$,
\begin{equation}
  T=\frac2c\,(\gamma-\gamma_I)(\gamma-\gamma_K),
\end{equation}
which for $\sqrt\delta\gg\epsilon$ becomes
$T\simeq(2/c)[(\gamma+\frac{a\epsilon+\epsilon'}{2})^2-\delta]$ and reduces to
the tree formula at $\epsilon=0$. For $\delta<0$ there is no second real
solution, and we choose $T=(2/c)(\gamma-\gamma_J)(\gamma-\gamma_J+a\epsilon)$,
with $\gamma_J$ the invariant line; since $\gamma_I=0$ and
$\gamma_K=-a\epsilon$ at $\delta=0$, this joins the form for $\delta>0$
continuously.

The function $\Phi$ is fixed by requiring that the tree relation
$\beta=-2\gamma$ of Eq.~\eqref{eq:D1beta} holds exactly along the
one-parameter curve. There $T=0$, so $D=\epsilon d_f\Phi$ and
$\beta=d\ln\Phi/d\ell$; using $\gamma=d\ln\psi/d\ell-a\epsilon/2$ with
$\psi=I_n(\zeta)$ from Eq.~\eqref{eq:eps-manifold} gives
\begin{equation}
  \Phi(\delta)=\biggl[\frac{(\zeta/2)^n}{\Gamma(n+1)\,I_n(\zeta)}\biggr]^2,
  \qquad
  D_c=\mathcal D(0,0)=\frac{d_f}{d},
  \label{eq:eps-Phi}
\end{equation}
with $J_n$ in place of $I_n$ for $\delta<0$. For $a=b=0$ this is
$\Phi=1/I_0(\sqrt\delta/\epsilon)^2$. Since $(\zeta/2)^{-n}I_n(\zeta)$ is a power series in
$\zeta^2\propto\delta$, $1/\Phi$ is an entire function of $\delta$, and $D$ is analytic
across the transition. Deep in the localized phase, $\sqrt\delta\gg\epsilon$,
$\Phi\propto e^{-2\zeta}$ up to a power-law prefactor, and along the curve
$\beta\simeq\epsilon'\ln(D/D_c)$ up to logarithmic corrections, as found above. The value
of $D_c$ is thus the real-space fractal dimension of the critical support
divided by $d$, a property of the observable and not of the flow.

The exponent, instead, is read from the slope at $D_c$. Near the fixed
point, $(\zeta/2)^{-n}I_n(\zeta)\,\Gamma(n+1)=1+\zeta^2/[4(n+1)]+\dots$, so that
Eq.~\eqref{eq:eps-Phi} gives $D\simeq D_c\{1-\delta/[2\epsilon'^2(n+1)]\}$, and
Eq.~\eqref{eq:eps-manifold} gives $\gamma\simeq\delta/[2\epsilon'(n+1)]$, so that $\beta=-2\gamma\simeq-\delta/[\epsilon'(n+1)]$. The factors $n+1$ cancel in the ratio, and
\begin{equation}
  \frac{d\beta}{d\ln D}\bigg|_{D_c}=2\epsilon'=2\epsilon-b\epsilon^2,
\end{equation}
which is Eq.~\eqref{eq:eps-slope},
in agreement with Eq.~\eqref{eq:eps-nu-loops}; for $a=b=0$ the slope is $2\epsilon$ and
$\nu=\frac12$. This is the dashed tangent in Fig.~\ref{fig:epsflowloops}(b). The slope
in the plane $(\ln D,\beta)$ is the eigenvalue of $\delta$ itself: it does not
depend on $d_f$, nor on $a$. $D_c$ belongs to the observable, $\nu$ to the flow.
This is not a coincidence: since $\beta=d\ln D/d\ell$, near the fixed point
$\ln D-\ln D_c$ grows along the unstable manifold as $e^{\lambda_t\ell}$, so that
$\beta\simeq\lambda_t(\ln D-\ln D_c)$ for any observable that depends smoothly on
the couplings and has $D_c\neq0$; this is the same relation $1/\nu=d\beta/d\ln g$
at $g_c$ of the scaling theory of Ref.~\cite{AALR1979}.
On the tree, instead, $D_c=0$ and the invariant of Eq.~\eqref{eq:betaflow} at
$\delta=0$ gives $\beta=-\sqrt{2cD}$, with an infinite slope.

\section{Realization of conformal symmetry at the critical point}
\label{app:conformal}

In Section~\ref{sec:lin-ads} we showed that, on functions that do not depend on the boundary coordinate $x$, the linear operator of the tree is a function of the Laplacian of $\mathbb H_2$, Eq.~\eqref{eq:LH2}. The isometry group of $\mathbb H_2$ is $\mathrm{SL}(2,\mathbb R)$, the global conformal group of the boundary line. In this appendix we ask how much of this group acts on the generating function $\chi$ (Appendix~\ref{app:conf-sl2}), what role is played by $\eta$ (Appendix~\ref{app:conf-eta}), and how the merger of the two exponents at $W_c$ appears in this language (Appendix~\ref{app:conf-jordan}). The answer is that only the dilatation is realized non-trivially, that the rest of the group enters through the reflection $\mu\leftrightarrow1-\mu$, and that at $W_c$ the dilatation acts by a Jordan block of rank two: the critical point on the tree is a logarithmic conformal quantum mechanics, and its logarithm is the BKT running of Eq.~\eqref{eq:riccati}.

We use the notation of Section~\ref{sec:lin-flow}: $h(s)$ is the scale profile of Eq.~\eqref{eq:factor}, $\psi=e^{-s/2}h$, and $\gamma\equiv m-\frac12=\psi'/\psi$, as in Section~\ref{sec:largeD}. The radial coordinate of $\mathbb H_2$ is $z=\tau=e^s$.

\subsection{The action of $\mathrm{SL}(2,\mathbb R)$}
\label{app:conf-sl2}

\paragraph{Generators.} The isometries of $\mathbb H_2$ are the M\"obius maps $w\mapsto(aw+b)/(cw+d)$, $ad-bc=1$, of $w=x+iz$. As vector fields, their generators are
\begin{equation}
  P=\partial_x,\qquad D=x\partial_x+z\partial_z,\qquad K=(x^2-z^2)\partial_x+2xz\,\partial_z ,
  \label{eq:conf-gen}
\end{equation}
which generate translations, dilatations and special conformal transformations, respectively. Computing the commutators of the vector fields, e.g.\ $[D,P]f=D\partial_xf-\partial_x(Df)=-\partial_xf$, one finds
\begin{equation}
  [D,P]=-P,\qquad [D,K]=K,\qquad [P,K]=2D .
  \label{eq:conf-alg}
\end{equation}
The quadratic Casimir is the Laplacian,
\begin{equation}
  D^2-\tfrac12\{P,K\}=\nabla^2 ,
  \label{eq:conf-casimir}
\end{equation}
which therefore commutes with $P$, $D$ and $K$. At the boundary $z=0$ the generators reduce to $\partial_x$, $x\partial_x$ and $x^2\partial_x$, which are the Hamiltonian, the dilatation and the special conformal generator of the conformal quantum mechanics of de~Alfaro, Fubini and Furlan~\cite{deAlfaro1976}. A boundary operator of weight $\Delta$ corresponds to a bulk field with $\Delta(\Delta-1)=M^2$, as in Eq.~\eqref{eq:AdSdict}.

\paragraph{Action on the physical sector.} The physical functions do not depend on $x$, and on them the generators and the inversion act as follows.
\begin{itemize}
  \item \emph{Translations act trivially}: $P\chi=0$.
  \item \emph{The dilatation is realized non-trivially}: on $x$-independent functions $D=z\partial_z=\partial_s$, and $e^{uD}$ maps $f(z)$ to $f(e^uz)$, i.e.\ $\tau\to e^u\tau$. This is an exact symmetry of the linear recursion, Eq.~\eqref{eq:lin}: $I_i=R_i^2(\eta+\sum_jI_j)$ is homogeneous of degree one in $(\eta,\{I_j\})$, and the recursion for the $R$'s does not involve the $I$'s. Hence
  \begin{equation}
    (I,\eta)\to e^{-u}(I,\eta),\qquad R\to R,\qquad \tau\to e^{u}\tau
    \label{eq:conf-dil}
  \end{equation}
  maps solutions into solutions and leaves $\tau I$ invariant.
  \item \emph{Special conformal transformations leave the sector}: $Kf(z)=2xz\,f'(z)$ depends on $x$. This does not contradict $[K,\nabla^2]=0$: $K$ preserves each eigenspace of $\nabla^2$, but the $x$-independent functions are the kernel of $P$, and $K$ does not commute with $P$. Using Eq.~\eqref{eq:conf-alg}, for $Pf=0$
  \begin{equation}
    P(Kf)=[P,K]f+K(Pf)=2Df ,
    \label{eq:conf-PKf}
  \end{equation}
  which vanishes only if $Df=0$, i.e.\ for $f=\text{const}$. $D$ instead preserves the kernel of $P$, since $[D,P]=-P$.
  \item \emph{The Weyl reflection is realized.} The inversion $w\mapsto-1/w$ exchanges the boundary points $0$ and $\infty$, and maps $z^\mu$ to $(z/|w|^2)^\mu$, the bulk-to-boundary propagator from $x_0=0$. On the spectrum it acts as $\mu\to1-\mu$, since the representations with Casimir $\mu(1-\mu)$ and $(1-\mu)\mu$ coincide. Equivalently, the Knapp--Stein intertwiner, i.e.\ the shadow transform of CFT, relates $\Delta$ and $1-\Delta$. The AAT--MF symmetry $\Lambda(\mu)=\Lambda(1-\mu)$ is the statement that the transfer operator commutes with this reflection.
\end{itemize}
The tree therefore sees the dilatation and the Weyl reflection, and these two together fix the dependence of the operator on the Casimir, Eq.~\eqref{eq:LH2}. The rest of $\mathrm{SL}(2,\mathbb R)$ acts on the auxiliary coordinate $x$.

\paragraph{The Casimir on the physical sector.} On $x$-independent functions $KPf=0$, and $PKf=2Df$ by Eq.~\eqref{eq:conf-PKf}, so that $\frac12\{P,K\}$ acts as $D$. Eq.~\eqref{eq:conf-casimir} then gives $\nabla^2\to D^2-D$, i.e.
\begin{equation}
  -\nabla^2=D(1-D)=\partial_s(1-\partial_s),
\end{equation}
which is Eq.~\eqref{eq:lapH2}, now derived from the group: on the physical sector the Casimir is a function of $D$ alone.

\begin{figure}[t]
\centering
\includegraphics[width=\columnwidth]{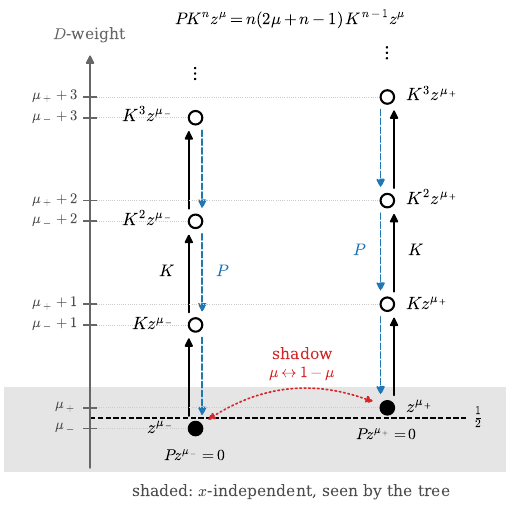}
\caption{The two lowest-weight towers of $\mathrm{SL}(2,\mathbb R)$ built on the primaries $z^{\mu_-}$ and $z^{\mu_+}$ (filled), with $Pz^{\mu_\pm}=0$, and their descendants $K^nz^{\mu_\pm}$ (open), of $D$-weight $\mu_\pm+n$. $K$ raises the weight by one (solid arrows) and $P$ lowers it (dashed), with the coefficient of Eq.~\eqref{eq:conf-PKn}. The two primaries are exchanged by the Weyl reflection $\mu\leftrightarrow1-\mu$, the shadow transform. Only the primaries are independent of $x$ (shaded), and only they enter the tree problem. Here $\delta=0.01$, i.e.\ $\mu_\pm=0.6$, $0.4$; at $W_c$ the two towers merge into a single logarithmic tower.}
\label{fig:sl2towers}
\end{figure}

\paragraph{Descendants.} Acting repeatedly with $K$ on $z^\mu$ gives
\begin{equation}
  Kz^\mu=2\mu\,x\,z^\mu,\qquad
  K^2z^\mu=2\mu\bigl[(1+2\mu)x^2-z^2\bigr]z^\mu ,
\end{equation}
and so on. All these functions are eigenfunctions of $-\nabla^2$ with the eigenvalue $\mu(1-\mu)$, as $[K,\nabla^2]=0$ requires. From $[D,K]=K$, $D(K^nz^\mu)=(\mu+n)K^nz^\mu$: $K$ raises the weight by one, and indeed $K^nz^\mu$ is homogeneous of degree $\mu+n$ in $(x,z)$. $P$ lowers it. Writing $PK^n=\sum_{k=0}^{n-1}K^k[P,K]K^{n-1-k}+K^nP$, and using $[P,K]=2D$, $Pz^\mu=0$ and the weights just found,
\begin{equation}
  \begin{aligned}
  PK^nz^\mu&=2\sum_{k=0}^{n-1}(\mu+n-1-k)\,K^{n-1}z^\mu\\
  &=n\,(2\mu+n-1)\,K^{n-1}z^\mu .
  \end{aligned}
  \label{eq:conf-PKn}
\end{equation}
For $n=1$ this is Eq.~\eqref{eq:conf-PKf}. Hence $z^\mu$ is the primary of a lowest-weight module: it is annihilated by $P$ and has $D$-weight $\mu$. The $K^nz^\mu$ are its descendants, with weights $\mu+n$. The two roots $\mu_\pm$ of $\Lambda(\mu)=1$ give two such towers, shown in Fig.~\ref{fig:sl2towers}. The roles of $P$ and $K$ are exchanged with respect to the usual conformal quantum mechanics because the source sits at $x_0=\infty$, which translations leave fixed; the inversion exchanges $P$ and $K$ and moves the source to $x_0=0$. By Eq.~\eqref{eq:conf-PKn}, null vectors occur only for $2\mu+n-1=0$, i.e.\ for $\mu=0,-\frac12,-1,\dots$, and there are none at $\mu=\frac12$.

At $W_c$ the logarithmic partner $\tilde{\mathcal O}=\partial_\mu z^\mu|_{1/2}$ of $\mathcal O=z^{1/2}$ (Appendix~\ref{app:conf-jordan}) has descendants $K^n\tilde{\mathcal O}=\partial_\mu(K^nz^\mu)|_{1/2}$. Differentiating Eq.~\eqref{eq:conf-PKn} with respect to $\mu$ at $\mu=\frac12$,
\begin{equation}
  PK^n\tilde{\mathcal O}=n^2K^{n-1}\tilde{\mathcal O}+2n\,K^{n-1}\mathcal O ,
\end{equation}
and on each pair $(K^n\mathcal O,K^n\tilde{\mathcal O})$ the dilatation acts by a Jordan block of weight $\frac12+n$. The whole module, and not only its primary, is therefore logarithmic.

The descendants have a geometric meaning. $z^\mu$ is the bulk-to-boundary propagator from $x_0=\infty$; $P$ translates the source, which is why $Pz^\mu=0$; $D$ rescales it; and $K$, which is $P$ conjugated by the inversion, moves the source away from $\infty$. Indeed, the orbit of $z^\mu$ is the family of propagators
\begin{equation}
  \Bigl[\frac{z}{z^2+(x-x_0)^2}\Bigr]^{\mu}=x_0^{-2\mu}\,\Bigl[z^\mu+\frac{Kz^\mu}{x_0}+O(x_0^{-2})\Bigr],
\end{equation}
all with Casimir $\mu(1-\mu)$, whose expansion in $1/x_0$ generates the $K^nz^\mu$ order by order. The descendants are the same solution, seen in a frame where its source is at a finite boundary point.

They are not, however, solutions of the tree problem. Eq.~\eqref{eq:LH2} was derived only on $x$-independent functions, where $-\nabla^2=\partial_s(1-\partial_s)$, and its extension to functions of $x$ is a choice: that $K^nz^\mu$ solves the extended equation says nothing about the cavity recursion. The tree selects the frame with the source at $x_0=\infty$, and in that frame only $D$ acts. Consistently, the descendants do not enter the observables. The moments of $I$ and the $D_q$ are Mellin transforms in $\tau$ of the $x$-independent $\chi$. For instance, for $0<q<1$, from $\int_0^\infty d\tau\,\tau^{-q-1}(1-e^{-\tau I})=\Gamma(1-q)I^q/q$,
\begin{equation}
  \langle I^q\rangle=\frac{q}{\Gamma(1-q)}\int_0^\infty\frac{d\tau}{\tau^{1+q}}\,\chi(\tau),
\end{equation}
and similar representations hold for the other $q$, Eq.~\eqref{eq:Dqb}. They probe only the dependence on $s$ of $x$-independent data, on which only $D$ acts. The operator content that reaches the observables is therefore the $2\times2$ block of $D$ on the primary pair:
\begin{itemize}
  \item the two roots $\mu_\pm$ of the same Casimir; the subdominant one sets the approach to the fixed point, e.g.\ the corrections $\propto\eta^{\mu_+-\mu_-}$ of Ref.~\cite{RTV2026lett}, and is the shadow of $\mu_+$, not a descendant;
  \item at $W_c$, the Jordan partner $\tilde{\mathcal O}=\tau^{1/2}\ln\tau$, which produces the logarithms; it is the generalized eigenvector of $D$ at the same weight $\frac12$, not a $K$-image, which would have weight $\frac32$;
  \item the nonlinear terms, e.g.\ $\tau^{2\mu}$ from $\chi^2$ beyond the wall, which are composites, not descendants.
\end{itemize}
In a one-dimensional CFT the descendants enter the correlation functions through the conformal blocks, when two boundary points are separated. Here every observable is a one-point function along the scale direction, with the source fixed at $x_0=\infty$, and there is no separation to expand in. The descendants might become physical at finite $d$, where the observables depend on position and the special conformal generator could act on a real spatial coordinate.

\paragraph{Representation content.} For a given value of the Casimir, $C=\mu_+\mu_-=\frac14-\delta$, the $x$-independent solutions form the two-dimensional space
\begin{equation}
  V_\delta=\Bigl\{h:\ \bigl(\partial_s-\tfrac12\bigr)^2h=\delta h\Bigr\},
\end{equation}
on which $D$ acts by a $2\times2$ matrix (Appendix~\ref{app:conf-jordan}). In terms of the spherical unitary dual of $\mathrm{SL}(2,\mathbb R)$, the localized phase ($0<\delta<\frac14$, $0<C<\frac14$) belongs to the complementary series, the delocalized phase ($\delta=-\kappa^2$, $C=\frac14+\kappa^2$) to the principal series, and $W_c$ ($C=\frac14$) is the point where the two series meet. Here unitarity refers to the representation of the group on functions on $\mathbb H_2$, not to a boundary theory, which, as usual in AdS, is not unitary for complex $\Delta$.

\subsection{The role of $\eta$}
\label{app:conf-eta}

The parameter $\eta$ does three things.

\emph{It breaks the hyperbolic symmetry.} In the supersymmetric $\sigma$-model~\cite{Efetov1983,Zirnbauer1986}, $\eta$ enters through the term $\eta\operatorname{Str}(\Lambda_{RA}Q)$, where $\Lambda_{RA}=\operatorname{diag}(1,-1)$ in retarded/advanced space (we write $\Lambda_{RA}$ to avoid confusion with the eigenvalue $\Lambda(\mu)$). This term breaks $\mathrm{U}(1,1|2)$ to $\mathrm{U}(1|1)\times\mathrm{U}(1|1)$, and its noncompact bosonic part from $\mathrm{SU}(1,1)\simeq\mathrm{SL}(2,\mathbb R)$ to $\mathrm{U}(1)$, the stabilizer of the point of $\mathbb H_2$ about which the order parameter is radial (Section~\ref{sec:lin-ads}). It acts like a magnetic field in a ferromagnet, with $\langle\operatorname{Str}\Lambda_{RA}Q\rangle\propto\rho(E)$, the density of states, which is not critical.

\emph{It is the only scale.} The dilatation~\eqref{eq:conf-dil} rescales $\eta$, which is the only parameter that breaks $D$. It has $D$-weight $-1$, like $I$, and enters the linear problem only through the invariant combination $\eta\tau=e^{s-\ell}$: in the source term of Eq.~\eqref{eq:chi}, $\tau R^2\eta=R^2e^{s-\ell}$, and at the nonlinear wall through $I/\eta$ and $b/\eta$ (Section~\ref{sec:lin-flow}). The region where Eq.~\eqref{eq:LH2} holds is bounded in the interior by the wall at $z_*=e^{\sigma}/\eta$, an infrared hard wall in the language of AdS. Since $\eta\sim1/N$, its radial position is $\ln z_*=\ln N+\text{const}$: the logarithm of the volume is a radial distance in AdS$_2$.

\emph{It turns the dilatation into the RG.} Lowering $\eta$ by a factor $e^{-u}$ moves the wall to $e^uz_*$, i.e.\ it acts with $e^{uD}$ on the position of the wall. The observables read the profile at the wall, $m(\ell)=m(s_*)$ with $s_*=\ell+\sigma$, so that $ds_*/d\ell=1$, and the unit rate is the normalization of $D$ in Eq.~\eqref{eq:conf-dil}. The $\eta$-RG is thus the holographic Wilsonian RG of Section~\ref{sec:lin-flow}, with the cutoff surface at the wall. For $m(s_*)$ to run, the profile must not move rigidly with the wall: the ratio $A:B$ of Eq.~\eqref{eq:hode} is fixed at an $\eta$-independent scale where the profile is anchored, as discussed in Section~\ref{sec:lin-flow} and shown in Fig.~\ref{fig:chiR}(b), while the wall fixes only the normalization $\chi_+$ and hence $c$, Eq.~\eqref{eq:cdef}.

\subsection{The Jordan block}
\label{app:conf-jordan}

\paragraph{The matrix of $D$.} On the space $V_\delta$ we choose the basis
\begin{equation}
  \begin{aligned}
  e_1^{(\delta)}&=\tfrac12\bigl(z^{\mu_+}+z^{\mu_-}\bigr)=z^{1/2}\cosh(\sqrt\delta\,s),\\
  e_2^{(\delta)}&=\frac{z^{\mu_+}-z^{\mu_-}}{\mu_+-\mu_-}=z^{1/2}\,\frac{\sinh(\sqrt\delta\,s)}{\sqrt\delta}.
  \end{aligned}
  \label{eq:conf-basis}
\end{equation}
Both are entire functions of $\delta$, and at $\delta=0$ they become $e_1=z^{1/2}$ and $e_2=z^{1/2}\ln z=s\,e^{s/2}$. Acting with $D=z\partial_z$, with $Dz^{\mu_\pm}=\mu_\pm z^{\mu_\pm}$ and $\mu_\pm=\frac12\pm\sqrt\delta$,
\begin{equation}
  \begin{aligned}
  De_1^{(\delta)}&=\tfrac12\bigl(\mu_+z^{\mu_+}+\mu_-z^{\mu_-}\bigr)
                 =\tfrac12e_1^{(\delta)}+\delta\,e_2^{(\delta)},\\
  De_2^{(\delta)}&=\frac{\mu_+z^{\mu_+}-\mu_-z^{\mu_-}}{2\sqrt\delta}
                 =\tfrac12e_2^{(\delta)}+e_1^{(\delta)} .
  \end{aligned}
\end{equation}
With the images of $e_1$ and $e_2$ as columns, the matrix of $D$ is
\begin{equation}
  D\big|_{V_\delta}=
  \begin{pmatrix}\frac12 & 1\\[2pt] \delta & \frac12\end{pmatrix}
  =\tfrac12\,\mathbb I+N_\delta,\qquad
  N_\delta=\begin{pmatrix}0&1\\ \delta&0\end{pmatrix}.
  \label{eq:conf-Dmatrix}
\end{equation}
It is analytic in $\delta$, and its eigenvalues are $\frac12\pm\sqrt\delta=\mu_\pm$. For $\delta>0$ it is diagonalizable with real eigenvalues, for $\delta<0$ it is diagonalizable over $\mathbb C$, with eigenvalues $\frac12\pm i\kappa$, and at $\delta=0$, i.e.\ at $W_c$, it is the Jordan block
\begin{equation}
  D\big|_{V_0}=\begin{pmatrix}\frac12&1\\0&\frac12\end{pmatrix},\qquad
  \bigl(D-\tfrac12\bigr)\,e_1=0,\qquad \bigl(D-\tfrac12\bigr)\,e_2=e_1 .
  \label{eq:conf-jordan}
\end{equation}
The Casimir, instead, stays diagonal: by Eq.~\eqref{eq:lapH2}, $-\nabla^2-\frac14=-(D-\frac12)^2$ on $x$-independent functions, which vanishes on all of $V_0$. Only the dilatation has the block, as $L_0$ does in two-dimensional logarithmic CFT~\cite{Gurarie1993}. Calling $\mathcal O=e_1$ and $\tilde{\mathcal O}=e_2$,
\begin{equation}
  \tilde{\mathcal O}=\lim_{\delta\to0}\frac{z^{\mu_+}-z^{\mu_-}}{\mu_+-\mu_-}=\partial_\mu z^{\mu}\big|_{\mu=1/2}.
\end{equation}
This is Cardy's mechanism for the appearance of logarithms~\cite{Cardy1999,Cardy2013}: the basis $z^{\mu_\pm}$ of eigenvectors becomes degenerate as $\delta\to0$, while the analytic basis~\eqref{eq:conf-basis} survives and carries the block.

\paragraph{Finite dilatations.} Since $N_\delta^2=\delta\,\mathbb I$, the exponential of Eq.~\eqref{eq:conf-Dmatrix} is
\begin{equation}
  \begin{aligned}
  e^{uD}\big|_{V_\delta}&=e^{u/2}
  \begin{pmatrix}\cosh\sqrt\delta u & \sinh(\sqrt\delta u)/\sqrt\delta\\
                 \sqrt\delta\,\sinh\sqrt\delta u & \cosh\sqrt\delta u\end{pmatrix}\\
  &\xrightarrow{\;\delta\to0\;}
  e^{u/2}\begin{pmatrix}1&u\\0&1\end{pmatrix}.
  \end{aligned}
  \label{eq:conf-finite}
\end{equation}
At $W_c$ a dilatation by $e^u$ thus maps $\tilde{\mathcal O}\to e^{u/2}(\tilde{\mathcal O}+u\,\mathcal O)$, and the logarithm is the term $u\,\mathcal O$. Eq.~\eqref{eq:conf-finite} is Eq.~\eqref{eq:hratio} written as a matrix.

\paragraph{Three phases, three conjugacy classes.} The traceless part $N_\delta$ of $D$ is an element of $\mathfrak{sl}(2,\mathbb R)$ with $\det N_\delta=-\delta$, and the sign of $\delta$ fixes its conjugacy class: hyperbolic for $\delta>0$ (localized), parabolic for $\delta=0$ (critical), elliptic for $\delta<0$ (delocalized). The critical point is the parabolic class, which is the one that cannot be diagonalized. This $\mathrm{SL}(2,\mathbb R)$ acts on the solution space $V_\delta$ and preserves the Wronskian: it is the transfer matrix of the radial evolution, not the isometry group of Appendix~\ref{app:conf-sl2}.

\paragraph{The Riccati flow is a M\"obius map.} With $Y=(\psi,\psi')^{\mathsf T}$, the equation $\psi''=\delta\psi$ reads $Y'=N_\delta Y$, so that $Y(s+u)=e^{uN_\delta}Y(s)$. Since $\gamma=\psi'/\psi$ is a projective coordinate on $Y$, it transforms by the M\"obius map of $e^{uN_\delta}$,
\begin{equation}
  \begin{aligned}
  \gamma(s+u)&=\frac{\sqrt\delta\sinh(\sqrt\delta u)+\gamma(s)\cosh(\sqrt\delta u)}
                   {\cosh(\sqrt\delta u)+\gamma(s)\sinh(\sqrt\delta u)/\sqrt\delta}\\
  &\xrightarrow{\;\delta\to0\;}
  \frac{\gamma(s)}{1+u\,\gamma(s)} .
  \end{aligned}
  \label{eq:conf-mobius}
\end{equation}
At $W_c$, therefore,
\begin{equation}
  \begin{gathered}
  \frac1{\gamma(s+u)}=\frac1{\gamma(s)}+u
  \quad\Longrightarrow\\
  \gamma=\frac1{s-s_0},\qquad m=\frac12+\frac1{s-s_0}:
  \end{gathered}
  \label{eq:conf-BKT}
\end{equation}
the BKT running is the parabolic M\"obius map, a translation of $1/\gamma$ at unit rate. Off criticality, Eq.~\eqref{eq:conf-mobius} gives the $\coth$ and $\tan$ solutions of Section~\ref{sec:lin-flow}. On the projective line, the physical branch ($\coth$, with $m$ decreasing from $+\infty$ to $\mu_+$) passes through $\gamma=\infty$ at $s=s_0$, where $h$ has a zero, i.e.\ $A$ and $B$ in Eq.~\eqref{eq:hode} have opposite signs; only the part $s>s_0$ is realized, since $\chi\ge0$.

\paragraph{Relation to the model of de~Alfaro, Fubini and Furlan.} The structure above is that of the conformal quantum mechanics of Ref.~\cite{deAlfaro1976}, with Hamiltonian $H=\frac12(p^2+g/x^2)$, in two precise senses. First, on $x$-independent functions $-\nabla^2=-z^2\partial_z^2$, so the Klein--Gordon equation $(-\nabla^2+M^2)h=0$ reads
\begin{equation}
  -\frac{d^2h}{dz^2}+\frac{M^2}{z^2}\,h=0 ,
\end{equation}
which is the zero-energy Schr\"odinger equation of that model, with the position replaced by $z=\tau$ and the coupling by $g=M^2=\delta-\frac14$. Its solutions $z^\nu$ have $\nu(\nu-1)=g$, i.e.\ $\nu=\mu_\pm$, and they are the non-normalizable zero-energy states discussed in Ref.~\cite{deAlfaro1976}. The localized phase corresponds to $-\frac14<g<0$, the critical point to $g=-\frac14$, where the two exponents merge, and the delocalized phase to $g<-\frac14$, the fall to the center of the inverse-square potential, which is the standard example of conformality lost~\cite{Kaplan2009}. Ref.~\cite{deAlfaro1976} considers only the repulsive case $g>0$. Second, in that model every generator $G=uH+vD+wK$ defines its own evolution time, $d\tau_G=dt/(u+vt+wt^2)$, and the generators fall into three classes according to the sign of $v^2-4uw$: hyperbolic ($v^2>4uw$, e.g.\ $D$), parabolic ($v^2=4uw$, e.g.\ $H$ and $K$) and compact ($v^2<4uw$). The corresponding times are a logarithm, an inverse and an arctangent of $t$, and only for compact generators does a finite range of $\tau_G$ cover the whole line $-\infty<t<\infty$. The Riccati equation~\eqref{eq:riccati}, $ds=d\gamma/(\delta-\gamma^2)$, is this relation with $t\to\gamma$, $u=\delta$, $v=0$ and $w=-1$, so that $v^2-4uw=4\delta$: the scale $s$ is the evolution time of a generator that is hyperbolic in the localized phase ($\coth$ solution), parabolic at $W_c$ (the solution $1/(s-s_0)$, in the class of the Hamiltonian), and compact in the delocalized phase ($\tan$ solution). The last property is the walking regime: $\gamma$ runs over the whole real line in the finite range $\Delta s=\pi/\kappa$. This is the classification of $N_\delta$ into conjugacy classes given above, seen from the side of Ref.~\cite{deAlfaro1976}.

\paragraph{The block is exact, and of rank two.} At $W_c$ we can use the exact Eq.~\eqref{eq:LH2} rather than its quadratic truncation. With $C=\mu(1-\mu)=\frac14-\varepsilon^2$ and $\Lambda\simeq1+\frac12\Lambda''\varepsilon^2$,
\begin{equation}
  \frac{d\tilde\Lambda}{dC}\Big|_{C=1/4}=-\tfrac12\Lambda''(\tfrac12)\ne0 ,
\end{equation}
so $1-\tilde\Lambda(C)$ has a simple zero at $C=\frac14$. The $x$-independent solutions belonging to this zero form the kernel of $-\nabla^2-\frac14=-(\partial_s-\frac12)^2$, which is spanned by $e^{s/2}$ and $s\,e^{s/2}$. The quartic and higher terms of $\Lambda(\partial_s)$ cannot split the double root in $\mu$, because they are functions of $C$ and the root is simple in $C$. Nor can they raise the rank: a term $s^2e^{s/2}$ would require $(-\nabla^2-\frac14)^2h=0$ with $(-\nabla^2-\frac14)h\neq0$, hence a double zero of $1-\tilde\Lambda$ in $C$, i.e.\ $\Lambda''(\frac12)=0$, which the strict convexity of $\ln\Lambda$ excludes. The rank two is forced by the Casimir being quadratic in $\mu$.

\paragraph{The dependence on $R$.} The exact modes are $e^{\mu s}f_\mu(R)$, Eq.~\eqref{eq:Tmu}, and at $W_c$ the pair is
\begin{equation}
  \begin{aligned}
  \mathcal O&=e^{s/2}f_{1/2}(R),\\
  \tilde{\mathcal O}&=\partial_\mu\bigl[e^{\mu s}f_\mu(R)\bigr]_{\mu=1/2}
  =e^{s/2}\bigl[s\,f_{1/2}(R)+\partial_\mu f_\mu(R)|_{1/2}\bigr].
  \end{aligned}
\end{equation}
The partner mixes with $\mathcal O$ through $\partial_\mu\ln f_\mu$. A general solution $\propto\tilde{\mathcal O}-s_0\,\mathcal O$ then gives $m(s\,|\,R)=\frac12+1/[s-s_0(R)]$, with unit residue in every $R$-bin and offset $s_0(R)=s_0-\partial_\mu\ln f_\mu(R)|_{1/2}$. This goes beyond the factorization of Eq.~\eqref{eq:factor} and could be tested with conditional data such as those of Fig.~\ref{fig:chiR}.

\paragraph{Amplitudes.} Let the profile be anchored at $s_w$ with Cauchy data $(h_w,h'_w)$, analytic in $\delta$. Then
\begin{equation}
  \begin{aligned}
  h(s)=e^{(s-s_w)/2}\Bigl[&h_w\cosh\sqrt\delta(s-s_w)\\
  &+\Bigl(h'_w-\frac{h_w}2\Bigr)\frac{\sinh\sqrt\delta(s-s_w)}{\sqrt\delta}\Bigr],
  \end{aligned}
\end{equation}
and comparing with $h=A\,e^{\mu_+s}+B\,e^{\mu_-s}$,
\begin{equation}
  \begin{aligned}
  A&=e^{-\mu_+s_w}\Bigl[\frac{h_w}2+\frac{h'_w-h_w/2}{2\sqrt\delta}\Bigr],\\
  B&=e^{-\mu_-s_w}\Bigl[\frac{h_w}2-\frac{h'_w-h_w/2}{2\sqrt\delta}\Bigr].
  \end{aligned}
\end{equation}
The amplitudes of the two colliding modes diverge as $\pm1/\sqrt\delta$, with opposite signs, while $h$ itself is analytic in $\delta$; on the delocalized side they become a complex-conjugate pair with $|A|\propto1/\kappa$. This is Cardy's observation~\cite{Cardy1999,Cardy2013} that the correlation functions are analytic in the parameter, while their decomposition into scaling pieces is not.

\paragraph{A second pair.} At $W_c$, $D_1\to0$ and, by Eq.~\eqref{eq:D1b}, $\langle\ln b\rangle=-\ell+\text{const}$. Under $\ell\to\ell+u$, $\ln b\to\ln b-u\,\mathbb I$, so $(\mathbb I,\ln b)$ is a Jordan pair of weight $0$ under the $\eta$-dilatation, the counterpart of $(1,\ln z)$ for a massless field. It is not a root of $\Lambda(\mu)=1$, since $\Lambda(0)=K$, and it concerns the bulk of the distribution of $I$, not its tail.

\section{Population dynamics: runs and detailed results}
\label{app:popdyn}

This appendix collects the details behind Section~\ref{sec:pd-results}: the runs and their error analysis, the fits of the flow of $D_1$ with their systematics, and the analysis of $D_2$.

\subsection{Runs and error analysis}
\label{app:pd-runs}

The code is C++ with OpenMP. The data of Fig.~\ref{fig:pd-prod} and of
Appendix~\ref{app:pd-fits} come from 100 independent populations of $N=10^7$
members, at $K=2$ and $E=0$, covering 28 disorders from $W=12$ to $30$:
eight populations at each of $W=18.1$, $18.17$, $18.2$; six at $W=18$ and
$18.3$; five at each of $W=16.5$, $17$, $17.25$, $17.5$; four at each of
$W=17.75$, $18.5$, $19$; and two at each of the other disorders
($W=12$--$16$ in steps of $0.5$, and $W=19.5$, $20$, $21$, $22$, $24$, $26$,
$30$). Each population evolves the whole grid of $\eta$, four points per
decade starting at $\eta=10^{-1/2}$, with common random numbers: all values
of $\eta$ use the same random numbers at every generation, so that the
population noise is strongly correlated in $\ell$ and largely cancels in
the derivatives of $S(\ell)$. The grid ends at $\eta=10^{-12.75}$
($\ell=29.4$) for $16.5\le W\le19$, at $10^{-14}$ for $W=16$, and at
$10^{-10}$ otherwise. The populations at $16.5\le W\le19$ have 1500
generations of burn-in and 2500 measured generations, the others 500 and
1800. The first $10\%$ of the measured generations are also discarded,
leaving 2250 or 1620 generations per population.

$S(\ell)$ is the mean over generations of $\langle w\ln I\rangle/\langle
w\rangle$, and $D_1=1-S'$ and $\beta=-S''/D_1$ come from the same weighted
local quadratic fit in $\ell$, over five points for $D_1(\ell)$
[Fig.~\ref{fig:pd-prod}(a)] and over seven for the flow
[Fig.~\ref{fig:pd-prod}(b)]; the last centered seven-point derivative is at
$\ell=27.63$. The statistical errors are the larger of two jackknife
estimates, one over the independent populations and one over the two halves
of each retained history. The second is needed where only two populations
are available.

\paragraph{Equilibration.}
Comparing the late and early halves of each history tests for residual
drift. At $W=17.5$ and $\ell=27.63$, on the delocalized arc, the late-half
estimate of $\beta$ exceeds the early-half one by $0.059(13)$, so the
populations are not fully equilibrated there at the largest $\ell$.

\paragraph{Finite population.}
The dependence on $N$ was checked with earlier runs (Appendix~\ref{app:pd-earlier}).
At $W=15$ and $18.17$, $N=10^6$ and $N=10^7$ agree within two standard
deviations at every $\ell$, while $N=10^5$ deviates in the crossover and at
large $\ell$. At $W=18.0$--$18.3$, the $N=10^7$ values of $S$ agree with the
$N=10^6$ ones, with a mean pull below one standard deviation in every window
of $\ell$ up to $\ell=28$. Finite-$N$ effects at $N=10^7$ are therefore
below the statistical errors of those runs; they have not been tested
directly at the precision of the present data.

\subsection{Fits of the flow of $D_1$}
\label{sec:pd-D1}
\label{app:pd-fits}

\paragraph{Integrated flow.}
Integrating the flow of Section~\ref{sec:lin-flow},
$dm/d\ell=-[(m-\frac12)^2-\delta]$ with $D_1=(2/c)[(m-\frac12)^2-\delta]$ and
$\beta=1-2m$, gives on both sides of the transition
\begin{equation}
  \begin{gathered}
  D_1=\frac{2\delta}{c\,\sinh^2\!\bigl[\sqrt\delta\,(\ell+\ell_0)\bigr]},\\
  S(\ell)=S_0+\ell+\frac{2\sqrt\delta}{c}\coth\bigl[\sqrt\delta\,(\ell+\ell_0)\bigr],
  \end{gathered}
  \label{eq:pd-S}
\end{equation}
continued to $\delta<0$ ($\sinh\to\sin$, $\coth\to\cot$) and to
$D_1=2/[c(\ell+\ell_0)^2]$ at $\delta=0$.
The same form, including its continuation to $\delta<0$, was obtained independently in Ref.~\cite{RTV2026prb} [their Eq.~(107)].

\paragraph{Critical bracket.}
We fit $S(\ell)$ directly, without finite differences, for the nine
disorders $17.5\le W\le19$ and $9.2\le\ell\le23.1$, with $c$ and $\ell_0$
common to all $W$, a constant $S_0(W)$ for each, $\delta=a(W-W_c)$, and $W_c$
free. This gives
\begin{equation}
  \begin{gathered}
  W_c=18.099,\qquad c=0.381,\\
  \ell_0=1.29,\qquad a=0.00960 .
  \end{gathered}
  \label{eq:pd-fit}
\end{equation}
The jackknife error on $W_c$, about $0.003$, is much smaller than the spread
over the tested windows in $\ell$, brackets of disorder and weighting
choices, $W_c=18.07$--$18.13$, $c=0.37$--$0.40$ and $a=0.0094$--$0.0099$,
which we take as the uncertainty. $W_c$ and $c$ move together: restricting
the fit to $17.75\le W\le18.5$ and $11.5\le\ell\le23.1$, for instance, gives
$W_c=18.132$ and $c=0.404$. The sum of the squared residuals, in units of
the pointwise errors, is $3189$ for $212$ nominal degrees of freedom.
Because the common random numbers correlate the points in $\ell$, this
number is not distributed as a $\chi^2$, and we use it only as a measure of
the mismatch: at this precision the data resolve departures from the
leading-order flow, Eq.~\eqref{eq:pd-S}. Adding a quadratic term to
$\delta(W)$ reduces it to $457$ for $211$ nominal degrees of freedom, with
$W_c$ still near $18.10$ in the same window. For these reasons we do not
attach a significance to the difference from $W_c=18.17$, the value quoted
for $K=2$ in Ref.~\cite{Vanoni2024}, which lies just above the range of our
fits. At $W=18.1$ the critical law holds up to the largest $\ell$ reached: at
$\ell=27.63$ the data give $D_1=0.00657(40)$ and $\beta=-0.051(26)$, against
$0.00628$ and $-0.069$ from Eq.~\eqref{eq:pd-fit}.

\paragraph{Delocalized side.}
The minimum of $D_1$ is located for 13 values of $W$ between $12$ and
$17.5$ by a quartic fit of $S(\ell)$ around it; Table~\ref{tab:pd-turning} lists a subset of them.
The flow predicts $D_A=2|\delta|/c$ and, with the same $\ell_0$ as at
criticality, the position of the minimum, $\ell_{\min}=\pi/(2\kappa)-\ell_0$
with $\kappa=\sqrt{a(W_c-W)}$; with the parameters of
Eq.~\eqref{eq:pd-fit} neither has a free parameter. The same predictions for the minimum, $D_A$ and $\ln N_A\simeq\pi/\kappa$, were obtained independently in Ref.~\cite{RTV2026prb}. Close to the transition, both agree with the data to a few percent: $D_A=0.0554$ and $0.0806$ against the measured $0.0565$ and $0.0844$ at $W=17$ and $16.5$, and $\ell_{\min}=7.82$, $11.39$, $14.00$ against $7.64$, $11.35$, $13.89$ at $W=15$, $16.5$, $17$. The agreement deteriorates where $D_A$ is no longer small: at $W=12$, $\ell_{\min}=5.20$ against $4.51$. The quadratic guide
$D_A\simeq0.051\,x+0.0029\,x^2$, $x=18.02-W$, of Fig.~\ref{fig:pd-prod}(c)
has a linear coefficient in agreement with $2a/c=0.050$. Its residuals,
however, are larger than the statistical errors, and its extrapolated zero
moves from $18.02$ to $18.10$ with equal weights, and to $18.15$ when the
weighted fit is restricted to $W\ge15$; the fit of the critical bracket
gives the more stable estimate of $W_c$. The curvature at the minimum,
$D_1''/D_A^2$, estimates $c$ through the parabola $\beta^2=2c(D_1-D_A)$. It
rises towards the transition, from $0.20$ at $W=12$ to $0.32$ at $W=15$,
$0.39$ at $W=16.5$ and $0.43$ at $W=17$.

\begin{table}[t]
\caption{Minimum $D_A$ of $D_1(\ell)$ and its position $\ell_{\min}$ on the
delocalized side, from quartic fits of $S(\ell)$, compared with the flow
predictions $2|\delta|/c$ and $\pi/(2\kappa)-\ell_0$,
$\kappa=\sqrt{a(W_c-W)}$, with the parameters of
Eq.~\eqref{eq:pd-fit}. See also Ref.~\cite{RTV2026prb}.}
\label{tab:pd-turning}
\begin{ruledtabular}
\begin{tabular}{ccccc}
\multicolumn{1}{c}{$W$} & \multicolumn{1}{c}{$D_A$} & \multicolumn{1}{c}{$2|\delta|/c$} & \multicolumn{1}{c}{$\ell_{\min}$} & \multicolumn{1}{c}{$\pi/(2\kappa)-\ell_0$}\\
\colrule
12   & 0.4116 & 0.3074 & 4.51  & 5.20 \\
13   & 0.3297 & 0.2570 & 5.29  & 5.81 \\
14   & 0.2519 & 0.2066 & 6.28  & 6.63 \\
15   & 0.1797 & 0.1562 & 7.64  & 7.82 \\
16   & 0.1142 & 0.1058 & 9.72  & 9.78 \\
16.5 & 0.0844 & 0.0806 & 11.35 & 11.39 \\
17   & 0.0565 & 0.0554 & 13.89 & 14.00 \\
17.5 & 0.0310 & 0.0302 & 18.8  & 19.4 \\
\end{tabular}
\end{ruledtabular}
\end{table}

\paragraph{Localized side.}
For each $W>W_c$ we fit Eq.~\eqref{eq:pd-S} separately
($6.9\le\ell\le23.1$), with $\alpha=2\sqrt\delta$, $c$ and $\ell_0$ free
(Table~\ref{tab:pd-alpha}). Close to the transition the rate is near
$2\sqrt{a(W-W_c)}$ with the parameters of Eq.~\eqref{eq:pd-fit}:
$\alpha=0.183$ against $0.186$ at $W=19$, and $0.258$ against $0.270$ at
$W=20$. The ratio $\alpha(W{=}20)/\alpha(W{=}19)$ is $1.408(8)$ statistically and
ranges from $1.397$ to $1.408$ across the tested lower cuts of the fit
window; it compares with $1.45$ for a square root and $2.1$
for a linear law. The decay rate of $D_1$ thus vanishes approximately as
$(W-W_c)^{1/2}$, while the localization length has $\nu_{\rm loc}=1$. The
departure from the leading-order flow grows with $W$: at $W=30$ the rate is
$0.525$, against $0.676$ extrapolated from Eq.~\eqref{eq:pd-fit}, and
$\alpha^2/4$ bends below the line $a(W-W_c)$
[Fig.~\ref{fig:pd-prod}(d)]; a weighted linear fit of $\alpha^2/4$ for
$18.5\le W\le20$ gives a slope $0.0086$ and a zero at $W=18.06$. Along each localized trajectory the invariant
$\beta^2/2-cD_1=\alpha^2/2$ gives effective values $c=0.34$--$0.43$ for
$18.5\le W\le19.5$, depending on the windows used for the derivatives and
the fit.

\begin{table}[t]
\caption{Localized decay rates $\alpha$ from fits of Eq.~\eqref{eq:pd-S}
over $6.9\le\ell\le23.1$ (statistical errors in the last digits), and the
prediction $2\sqrt{a(W-W_c)}$ with the parameters of Eq.~\eqref{eq:pd-fit}.}
\label{tab:pd-alpha}
\begin{ruledtabular}
\begin{tabular}{ccc}
\multicolumn{1}{c}{$W$} & $\alpha$ & \multicolumn{1}{c}{$2\sqrt{a(W-W_c)}$}\\
\colrule
18.5 & 0.1219(9)  & 0.124 \\
19   & 0.1832(9)  & 0.186 \\
19.5 & 0.2226(39) & 0.232 \\
20   & 0.2579(8)  & 0.270 \\
21   & 0.3118(23) & 0.334 \\
22   & 0.3518(4)  & 0.387 \\
24   & 0.4129(19) & 0.476 \\
26   & 0.4601(8)  & 0.551 \\
30   & 0.5247(19) & 0.676 \\
\end{tabular}
\end{ruledtabular}
\end{table}

\paragraph{The constant $c$.}
$c$ is a parameter of the flow, not predicted by it. Three determinations
use largely independent data: the common fit of the critical bracket,
$c\simeq0.38$ ($0.37$--$0.40$); the invariant along localized trajectories,
$0.34$--$0.43$ for $18.5\le W\le19.5$; and the curvature at the delocalized
minimum, which reaches $0.39$--$0.43$ for $16.5\le W\le17$. They are
consistent, so near the transition $c\simeq0.4$ is a property of the flow.
It is not constant along the delocalized arc: the curvature estimate falls
to $0.20$ at $W=12$. Under a rescaling $\ell\to k\ell$,
$c\to c/k^2$, so the comparison with the RRG assumes that
$\ln(1/\eta)$ and $\ln N$ match with unit slope.
The equality of the finite-$\eta$ and finite-$N$ correction exponents that underlies this assumption was shown in Ref.~\cite{RTV2026lett}; the amplitudes, and hence $c$, can differ between the two.

\paragraph{Approach to the ergodic point.}
Near $D_1=1$ the flow is $\beta\simeq\lambda(1-D_1)$, with predicted slope
$\lambda=1$. Fitting $S(\ell)$ at $W=13$ and $14$ with the next order
included, $1-D_1=Ae^{-\lambda\ell}+Be^{-2\lambda\ell}$, gives
$\lambda=0.993$--$0.999$ when the fit starts at $D_1\simeq0.95$. Windows
that start farther from $D_1=1$ return smaller rates, especially at $W=14$,
so the fitted rate approaches $\lambda=1$ as the window is restricted to the
neighborhood of the ergodic point.

\subsection{The flowing $D_2$}
\label{sec:pd-D2}
\label{app:pd-D2}

Our raw
data agree with those of Ref.~\cite{TM2019}: $2\eta\langle I^2\rangle=0.071$ at
$\log_{10}(1/\eta)=2$ and $0.058$ at $7.25$, and the fit with $\zeta=\frac12$ over
their range at $W=18$ gives $c_1=0.043$, against their $0.041$. The derivative,
however, is not described by $\zeta=\frac12$ [Fig.~\ref{fig:pd-D2}(a)]. With
$\zeta$ free, the fit gives $\zeta=0.82$--$0.85$ for $4.6\le\ell\le16.1$ and
$0.90$--$1.07$ for $12\le\ell\le29.4$ at $W=18.1$--$18.2$. A rising effective
$\zeta$ is what a form without offset returns when fitted to
$\text{const}+A/(\ell+\ell_0)$: $\zeta_{\rm eff}=\ell/(\ell+\ell_0)$. The direct
fit of the latter on $9.2\le\ell\le29.4$ gives $A=1.89(1)$ and
$\ell_0=1.59(4)$ at $W=18.1$, and describes $D_2$ over the whole range
[Fig.~\ref{fig:pd-D2}(a)]. The same conclusion, $\zeta=1$ with the smaller effective exponent of Ref.~\cite{TM2019} explained as preasymptotic, was reached independently in Ref.~\cite{RTV2026prb} (Sec.~VI~A).

The ratio $D_2/D_1$ is nearly constant along each trajectory with
$D_1\ll1$: on the
localized side it stays within $0.320$--$0.328$ at $W=19$, $0.304$--$0.308$ at
$W=20$ and $0.290$--$0.293$ at $W=21$, while $D_1$ falls by a factor $15$ to
$20$; near $W_c$ it drifts slowly, from $0.335$ to $0.354$ over
$6\le\ell\le28$ at $W=18.1$, and flattens at large $\ell$
[Fig.~\ref{fig:pd-D2}(b)], as expected from the $O(1/\ell)$ correction. On
the delocalized arc the ratio rises towards one, as it must since all
$D_q\to1$ at the ergodic point (from $0.36$ to $0.43$ at $W=17$, and from
$0.38$ to $0.84$ at $W=16$, for $6\le\ell\le23$).

\subsection{Earlier runs}
\label{app:pd-earlier}

The conditional generating function of Fig.~\ref{fig:chiR} and the checks
of finite-$N$ effects use earlier runs, in which each value of $\eta$ was evolved with independent
random numbers. The main set used $N=10^6$, four values of $\eta$ per decade
from $10^{-2}$, 80 generations of burn-in and 1500 measured generations per
$\eta$ (the first $10\%$ of which were also discarded), and the same 28
disorders; thirteen of them, between $W=16$ and $19$, extend to
$\eta=10^{-14}$ ($\ell\le32$). Errors were computed from the integrated
autocorrelation time and inflated by factors $1.25$ to $1.8$, as indicated
by comparing the two halves of each run. A second set at $N=10^7$ covered
$W=18.0$, $18.1$, $18.2$ ($9.2\le\ell\le27.6$), the delocalized arcs at
$W=16$, $16.5$ and $17$, and checks at $W=17$, $18.17$ and $18.3$.

\bibliography{references}

\end{document}